\documentclass[aps,prd,reprint,nofootinbib]{revtex4-2}
\usepackage{xcolor}
\usepackage{amsmath}
\usepackage{tikz}
\usepackage{xspace}
\usetikzlibrary{calc}
\usepackage{orcidlink}
\def\apjs{ApJS}
\newcommand{\chieff}{\ensuremath{\chi_{\rm eff}}\xspace} 
\newcommand{\Mc}{\ensuremath{\mathcal{M}_c}\xspace} 
\newcommand{\detone}{\ensuremath{I}\xspace} 
\newcommand{\dettwo}{\ensuremath{J}\xspace} 
\newcommand{\score}{\texttt{SCoRe}\xspace} 
\newcommand{\posterior}{\ensuremath{P}\xspace} 
\newcommand{\prior}{\ensuremath{\pi}\xspace} 
\newcommand{\likelihood}{\ensuremath{\mathcal{L}}\xspace} 
\newcommand{\cut}{\ensuremath{l_\text{UV}\xspace}} 

\newcommand{\influenceAlpha}{\ensuremath{10^{-3}}\xspace}
\newcommand{\influenceDHigh}{\ensuremath{4}\xspace}
\newcommand{\influenceNEventsHigh}{\ensuremath{10^{10}}}
\newcommand{\influenceDLow}{\ensuremath{\frac{5}{6}}\xspace}
\newcommand{\influenceNEventsLow}{\ensuremath{3 \times 10^{7}}\xspace}
\newcommand{\influenceDeltaM}{\ensuremath{2}\xspace}
\newcommand{\exampleEventMc}{\ensuremath{8.15M_{\odot}}\xspace}
\newcommand{\exampleEventdL}{\ensuremath{1.19}Mpc}
\newcommand{\exampleEventSmallBeta}{\ensuremath{1.16}}
\newcommand{\exampleEventLargeBeta}{\ensuremath{35.3}}

\newcommand{\thresholdSNR}{\ensuremath{8}\xspace}

\graphicspath{{./figures/}}

\begin{document}

\title{Detecting Unmodeled, Source-Dependent Signals in Gravitational Waves with \score}

\date{\today}

\author{Guillaume Dideron \orcidlink{0000-0002-5222-7974}}\email{gdideron@perimeterinstitute.ca}
\affiliation{Department of Physics and Astronomy, University of 
Waterloo, 200 University Avenue West, Waterloo, ON N2L 3G1, Canada}
\affiliation{Perimeter Institute for Theoretical Physics, 31 Caroline Street North, Waterloo, ON N2L 2Y5, Canada}

\author{Suvodip Mukherjee \orcidlink{0000-0002-3373-5236}}\email{suvodip@tifr.res.in}
\affiliation{Department of Astronomy \& Astrophysics, Tata Institute of Fundamental Research, 1, Homi Bhabha Road, Colaba, Mumbai 400005, India}

\author{Luis Lehner \orcidlink{0000-0001-9682-3383}}\email{llehner@perimeterinstitute.ca}
\affiliation{Perimeter Institute for Theoretical Physics, 31 Caroline Street North, Waterloo, ON N2L 2Y5, Canada}

\begin{abstract}
New physics and systematic errors can lead to deviations between the models
used to analyze gravitational wave data and the actual signal. 
Such deviations will generally be correlated between detectors and manifest
differently across the gravitational wave source parameter space.
The previously introduced \score framework uses these features to distinguish
these deviations from noise and extract physical information from their
source-dependent variation. 
In this work, we further analyze the hierarchical component of the method---we include
the expected dependence of the deviations on the source parameters into the
inference process, obtaining more physically informative results.
As a specific example, we study a deviation that scales as a
power law of the mass scale of black hole binaries---as, for example, in Effective
Field Theory of gravity.
We show how the signal-to-noise ratio of the cross-correlated residual power can be used to recover 
the power-law index.
We demonstrate how both the distribution in source and deviation strength
determine which region of source parameter space influences the inference most.
Finally, we forecast the constraint on the power law index
for a network of two Cosmic Explorer-like detectors with a year of observation period.
\end{abstract}

\maketitle

\section{Introduction}

Gravitational Waves (GWs) give us a window to an otherwise invisible side of
the universe.  The first three observation runs of Advanced
LIGO~\cite{theligoscientificcollaborationAdvancedLIGO2015}, Advanced Virgo~\cite{acerneseAdvancedVirgoSecondgeneration2015}, and KAGRA~\cite{somiyaDetectorConfigurationKAGRA2012,thekagracollaborationInterferometerDesignKAGRA2013,akutsuOverviewKAGRADetector2021} have allowed the study
of a wide range of phenomena, in
cosmology,
astrophysics, and fundamental
physics (see, e.g.~\cite{theligoscientificcollaborationAdvancedLIGO2015,
acerneseAdvancedVirgoSecondgeneration2015,
abbottConstraintsCosmicExpansion2023a,abbottPopulationMergingCompact2023,
abbottPopulationMergingCompact2023,
theligoscientificcollaborationTestsGeneralRelativity2021a,
Yunes:2016jcc,
Capano:2019eae,
Mandel:2021smh,
Godzieba:2020tjn}).  
At the core of these efforts, the inference of physical information from GWs 
relies on our understanding of the emission mechanism, the sources of GWs,
and the noise in the detectors. These assumptions we dub the Standard Model
(SM) of GWs.
The SM is used to compute waveform models of the signal from a target 
source~\footnote{Note that the detection of GWs itself does not require models. For
	example, the coherent
	WaveBurst~\cite{klimenkoConstraintLikelihoodAnalysis2005,klimenkoConstraintLikelihoodMethod2006,klimenkoLocalizationGravitationalWave2011,klimenkoMethodDetectionReconstruction2016,klimenkoWaveletMethodDetection2004}
	pipeline identifies excess power that is coherent across detectors and
	has been used in LVK analyses, and the
	\textsc{MLy}~\cite{sklirisRealTimeDetectionUnmodelled2020} pipeline
	uses convolution neural networks to identify coherent signals between
	detectors.
}.
These waveform models generally have finite precision, and may not fully implement
known physics (such as overlapping signals, eccentricity, dynamical tides,
etc), and most omit yet unknown physics (such as deviations from General
Relativity (GR) or alternative compact objects). Such mismodeling gives rise to a
difference between the physical signal in the data and the SM waveforms. We
term such differences Beyond Modeled (BM) signatures. The presence of new physics
in the data~\cite{theligoscientificcollaborationTestsGeneralRelativity2021a} as
well as the impact of systematic errors on parameter
estimation~\cite{abbottEffectsWaveformModel2017,chatziioannouPropertiesMassiveBinary2019}
and on the detection of new physics~\cite{guptaPossibleCausesFalse2024a} are regularly studied, 
and SM waveforms are continuously improved.

Next generation GW detectors (in the $\approx [1-10^4]$Hz window) such as Cosmic Explorer
(CE)~\cite{reitzeCosmicExplorerContribution2019} and the Einstein
Telescope~\cite{punturoEinsteinTelescopeThirdgeneration2010}, are expected to
detect events with maximum Signal-to-Noise Ratios (SNRs) on the order of
$\mathcal{O}\left(10^{3}\right)$ and at rates in the order of
$\mathcal{O}\left(10^{5}\right)$yr$^{-1}$~\cite{maggioreScienceCaseEinstein2020,evansHorizonStudyCosmic2021}.
As a comparison, the $\sim 90$ confident events currently in the third
Gravitational-Waves Transient Catalogs have maximum SNRs on the order
$\mathcal{O}\left(10^{1}\right)$~\cite{ligoscientificcollaborationandvirgocollaborationGWTC1GravitationalWaveTransient2019,ligoscientificcollaborationandvirgocollaborationGWTC2CompactBinary2021,ligoscientificcollaborationGWTC3CompactBinary2023,theligoscientificcollaborationandthevirgocollaborationGWTC21DeepExtended2024}.
This increase in performance means that BM
signatures will likely become more significant compared to random noise. Without a
proper grasp of their effect and origin, we may miss new physical effects or
create biases in our analyses~\cite{huAccumulatingErrorsTests2023}. Options to
address this shortcoming have been presented and improved in multiple techniques. 
For instance, BM signatures may be detected in a data-driven way through the residual 
test used, for example, in~\cite{abbottTestsGeneralRelativity2016,abbottTestsGeneralRelativity2019,abbottPropertiesAstrophysicalImplications2020a,theligoscientificcollaborationTestsGeneralRelativity2021a}.

An important feature of BM signatures is that they can vary across
events with the GW source properties such as mass, spin,
eccentricity, etcetera. 

We believe this point has so far not yet 
been fully exploited in the analysis of GW data.
This has been studied, for example, in~\cite{zimmermanCombiningInformationMultiple2019,isiHierarchicalTestGeneral2019,isiComparingBayesFactors2022a,payneFortifyingGravitationalwaveTests2023,mageeImpactSelectionBiases2024,payneCurvatureDependenceGravitationalwave2024,zhongMultidimensionalHierarchicalTests2024}.
In this work, we explore the inclusion of the source parameter dependence of BM
signatures in \score~\cite{dideronNewFrameworkStudy2023},  a method to detect and
extract BM signatures with minimal assumptions on their form by using the
residual power left after subtracting the model from the data. In brief, the
method has 3 steps: (1) The cross-correlated residual power in pairs of
detectors is computed to reduce the effect of uncorrelated noise. (2) That
power is filtered for physically motivated behavior. (3) Information from
different events is combined to inform on the origin of the BM signature. In
this work, after briefly reviewing the formalism underlying \score, we 
further elaborate on the method's
third step. We demonstrate that the dependence of a BM deviation on source parameters
can be recovered with the method, and forecast the constraints on SM deviations from GW
events by a two CE-like detector network over a year.
The residual test mentioned above uses the \emph{BayesWave} method to identify any
excess residual power in the detectors by fitting
wavelets data to coherent
signal~\cite{cornishBayeswaveBayesianInference2015,littenbergBayesianInferenceSpectral2015,
cornishBayesWaveAnalysisPipeline2021,BayesWaveBayesWave107}. 
Instead, the \score method takes the cross-correlation between detectors
and compliments the residual test by looking for specific morphology
in the residual power and/or the correlation of the residual with source properties.

This work is organized as follows: In
Sec.~\ref{sec:SCoRe formalism}, we describe the \score framework. Notably, how
the Cross-correlated Residual Power (CRPS) is constructed, how templates are
used to filter the CRPS for BM physics, and finally how the dependence of BM
signatures on source parameters are used to infer both the presence of BM
signature and general properties of its origin. In Sec.~\ref{sec:Toy model}, we
describe a specific type of possible BM signature: a phase shift in the
waveform of binary black holes due to a BM effect that scales as a power law of
the mass scale of the system. In Sec.~\ref{sub:CRPS scaling} and Sec. \ref{sub:influence}, we show how the SNR of
the CRPS can be used to recover the index of this power law when the BM
signature is small and how both the distribution in source and
deviation strength determines which region of source parameter space has the
most influence on the inference. Finally, in Sec.~\ref{sub:Constraints for CE-like network},
we forecast the constraint on the power law index that can be obtained with
two Cosmic Explorer-like detector network over a year. We conclude in Sec.~\ref{sec:Conclusions}.

\section{\texttt{SCoRe} Framework}
\label{sec:SCoRe formalism}

\subsection{Beyond Model deviations}%
\label{sub:Beyond Model deviations}

\begin{figure}[htpb]
	\centering
	\includegraphics[width=\columnwidth]{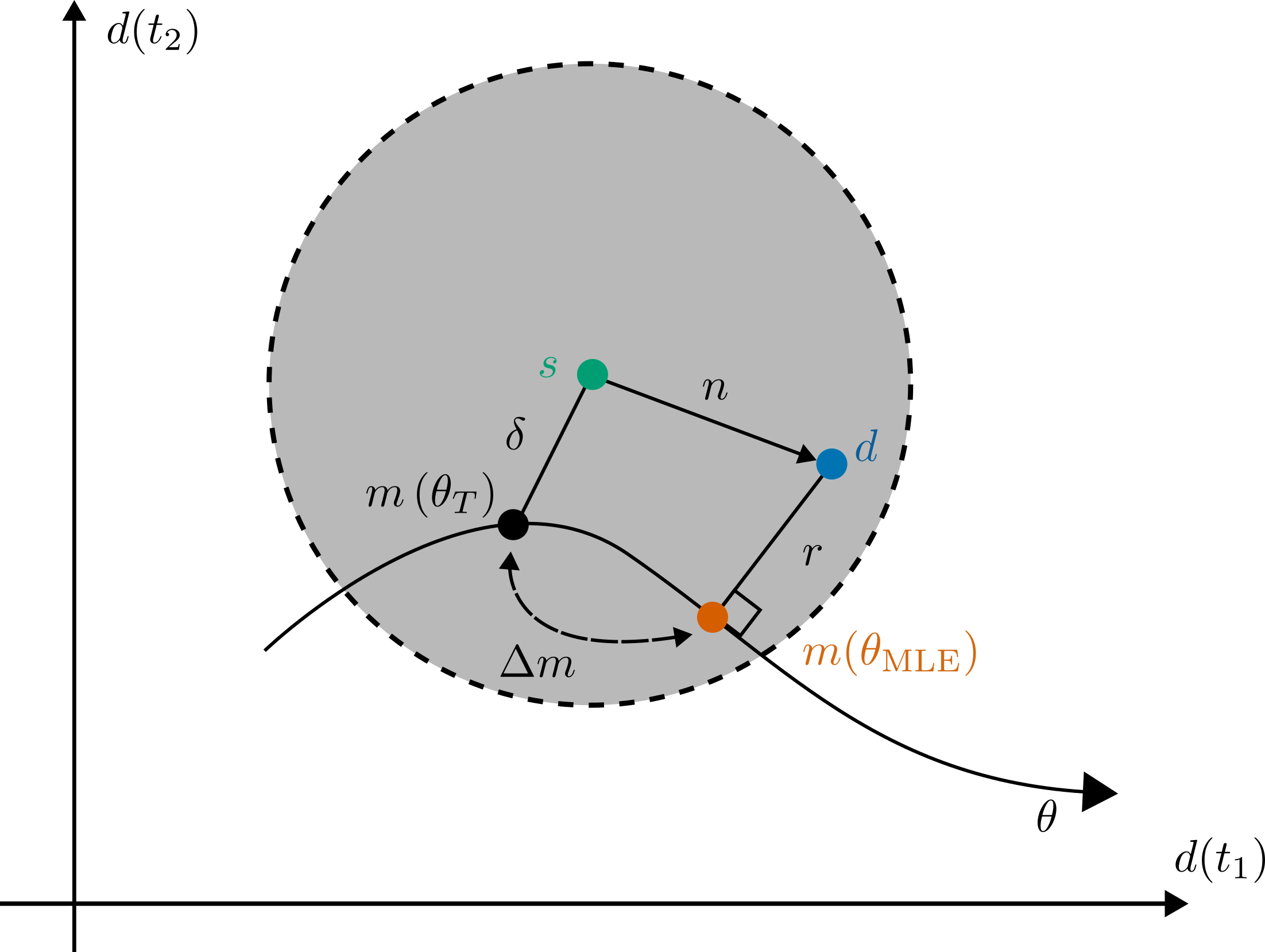}
	\caption{Geometric illustration of how the data $d$ arises from the
		signal $s$ and the noise $n$. If some
		unmodelled physics is present, the data converges to the BM signature  $\delta$ 
		in the absence of noise.
		Both the noise and the BM signature cause the MLE of the source parameters, 
        $\theta_\text{MLE}$,
		to differ from the true parameters $\theta_{T}$. 
		The residual $r$ is the difference between the data and the model.
		The shaded area represents the probability distribution of the noise 
        (where the noise may take the signal to the data with a certain level of 
        probability).
	}
	\label{fig:geometric_picture}
\end{figure}

In this section, we review the \score framework. We start by defining
the strain data $d$
\begin{align}
	\label{eq:def_strain}	
	d(t) = s(t) + n(t),
\end{align}
where $s(t)$ is the strain caused by the GW signal, while $n(t)$
is the noise in the detector. 
In the absence of noise and presence of a signal, the measured strain
$d(t)$ converges to the signal $s(t)$.
The residual data $r(t ; \theta)$ is obtained by subtracting a model $m \left(
\theta \right)$\footnote{It is important to note that in general the true signal in the data $s(t)$ can be unknown and differ from the model used in the analysis $m(t)$. This is discussed in more detail in the later part of this section.}, evaluated at source parameters $\theta$, from the data:
\begin{align}
	\label{eq:def_residual}
	r
	\left(t ; \theta \right)
	\equiv
	d(t)
	-
	m \left(t ; \theta \right).
\end{align}
The residual data contains noise, the difference between the model at $\theta$
and at the true parameters $\theta_{T}$, and the BM signature $\delta$:
\begin{align}
	\label{eq:def_residual}
	r
	\left( t ; \theta \right)
	=
	m \left( t ; \theta_\text{T}  \right) 
	-
	m \left( t ; \theta \right)
	+
	\delta ( t)
	+
	n (t)
	.
\end{align}
When using the Maximum Likelihood Estimator (MLE) $\theta=\theta_\text{MLE}$,
the bias $\Delta m (t) \equiv m \left( t ; \theta_\text{T} \right) - m \left( t ; \theta \right)$
is due to noise and any degeneracy between the BM signature and the model
\footnote{The bias $\Delta \theta^{i}$ on the set of parameters $\theta^{i}$,
	in the limit of high SNR, is given by 
$
\Delta \theta^{i}
=
\left( \Gamma^{-1} \right) ^{ij}
\left( \partial_{j} m \left( \theta_\text{MLE}   \right) | n + \delta \right)
$,
where $\Gamma$ is the Fisher matrix at the MLE, 
See
\cite{flanaganMeasuringGravitationalWaves1998,millerAccuracyRequirementsCalculation2005,cutlerBigBangObserver2006,cutlerLISADetectionsMassive2007}.
}.
This is shown in Fig.~\ref{fig:geometric_picture}, where both $\delta$ and $n$
cause the MLE to be different from the true parameters. In this work, we
define the BM signature as $s - m \left( \theta_{T}  \right)$. 
If one considers the bias due to degeneracy between the model and the BM
signature as part of the BM signature, then the BM signature is $\delta = s - m
\left( \theta_\text{MLE}   \right)$, and the formalism described below remains
the same. 

\subsection{Cross-correlated Residual Power}%
\label{sub:Cross-correlated Residual Power}

A BM signature will be correlated between detectors, whereas noise in different
detectors is largely not
\footnote{Some sources of noise are correlated between detectors. In particular,
	correlated noise due to magnetic coupling is expected to impact analysis 
	done with 3G detectors (Einstein Telescope) sensitivity~\cite{himemotoImpactCorrelatedMagnetic2017,himemotoCorrelatedMagneticNoise2019,janssensProspectsIsotropicGravitational2023a,himemotoDistinguishingStochasticGravitationalwave2023a,janssensCorrelated00140Hz2024}. Some of these works also suggest mitigation strategies. We do not concern ourselves with
 this issue at this time (for simplicity) but note that our framework could also be used to study correlated noise itself as a BM signature.}
.
This is the principle behind the first step of \texttt{SCoRe}.
On one hand, the mean time average of the product of BM signatures in
two detectors will be non-zero. On the other hand, the mean
time average of the product of noise in the detectors will converge to zero as
the timescale over which the average is taken increases. 

By using this aspect, we define the cross-correlated residual power
(CRPS) $D^{\detone \dettwo}$ as the mean of the product of the residuals
in a pair of detectors $\detone$ and $\dettwo$ over a timescale $\tau$:
\begin{align}
	D^{\detone \dettwo}(t ; \theta)
	&\equiv
	\left\langle
	r^{\detone} r^{\dettwo}
	\right\rangle
 \nonumber
 \\
 &=
	\frac{1}{\tau (t)}
	\int^{t + \frac{\tau(t)}{2}}_{t - \frac{\tau(t)}{2}}
	dt'
	r^{I}(t' ; \theta) r^{J}(t'+\Delta t; \theta),
	\label{eq:def_cross_correlation}
\end{align}
where $\Delta t$ is the time delay between detectors. (The value of  $\Delta t$ may 
be fixed or jointly inferred/marginalized over as part of $\theta$ in the analysis.)
The important feature of the CRPS is that it converges to
$\delta^{\detone}\delta^{\dettwo}$ as $\tau$ goes to infinity.
In analyzing GW waveforms, one usually uses the phase information of the signal.
This phase information is lost when taking the CRPS of the data.
In the \score framework, we instead focus on filtering the power emitted over
the timescale $\tau$. This is because we aim to remain agnostic about the specific
phase evolution of the BM signature, while still capturing any excess or dearth of energy loss 
that may indicate the presence of a deviation from the SM. Last, we stress
that one can still
filter out some expected time evolution behavior to test more specifically for 
certain BM signatures, as we explain in the next section.

\subsection{Cross-correlated residual power templates}%
\label{sub:Cross-correlated residual power templates}

The second step of \score filters for the expected time change in the CRPS
given a template time function $Z(t;\theta)$, which describes the expected time
evolution of the excess power due to the BM signature.
The templates $Z(t;\theta)$ are constructed to search for a physically possible
signal from the data in a model-independent (or dependent) way
\footnote
{
Examples discussed in~\cite{dideronNewFrameworkStudy2023} include templates looking 
for a chirp-like behaviour.
}.
We define the \emph{BM} SNR $\alpha$ for
the template $Z(t;\theta)$ as
\begin{align}
	\label{eq:def_power_projection}
	\alpha
	&=
	\int ^{t_{e}} _{t_s}
	d t
	W \left( t \right) 
	D^{IJ} \left( t ; \theta \right) 
	Z \left( t ; \theta \right) ,\\
	\text{where\,} W \left( t \right) 
	&\equiv
	\left(
		\int^{t_e}_{t_s}
		d t'
		N^{2}(t')
		Z^{2}(t')
	\right)^{-1}\nonumber
	.
\end{align}
In the above equation, $t_s$ and $t_e$ are the start and end times of the signal.
The CRPS noise $N(t)$ is the cross-correlated noise at time $t$--- the 
mean ensemble average of the signal in the absence of a BM signature.
The weighting function $W \left( t \right)$ is the inverse of the variance
of the filtered CRPS, so that $\alpha$ is the Wiener
filter~\cite{wienerExtrapolationInterpolationSmoothing1949} for the template.
Note that the match-filtering is performed in time, rather than in frequency.
We have assumed that $\tau$ is
greater than the timescale of auto-correlation of the noise.
Here, we have chosen the templates $Z(t;\theta)$ so that their time evolution
is solely determined by the SM parameters (while the amplitude/SNR of the
template is determined by the BM parameter). 
The CRPS data, $D^{IJ}$, depends on the SM parameters, since the residual 
is different for different source parameters.

\subsection{Modelling dependence on source parameters}%
\label{sub:Modelling dependence on source parameters}

The third step of \score combines the information from different events by
modeling how BM signatures in different events depend on the source parameters.
We introduce a set of hyperparameters $\Delta_{M}$ to describe this dependence across
the space of source parameters.
These hyperparameters can be inferred from a population of events using 
Hierarchical Bayesian inference, which we describe in Appendix~\ref{app:Hierarchical Bayesian Model}.
Given a set of observed data $\{d^{\detone}\}_i$ (pairs of data strain for each pair of detectors in the network)
for each event $i$, 
we marginalize over the source parameters
$\theta_{i}$ and the individual event BM signatures $\delta_{i}$ 
to obtain the posterior on $\Delta_M$:
\begin{align}
	\label{eq:hierarchical_posterior}
	\posterior 
	\left( \Delta_M | \{d^{\detone}\}_i \right)
	&\propto
	\prior \left( \Delta_M \right)
	\prod_{i}
	\int d \theta_{i}
	\likelihood
	\left( \{d^{\detone}\}_i | \theta_{i}, \Delta_M \right)
	\prior 
	\left( \theta_{i} \right),
\end{align}
where $\posterior (\Delta_M | \{d^{\detone}\}_i)$ is the posterior distribution
on the hyperparameters $\Delta_M$ given the data $\{d^{\detone}\}_i$;
$\prior (\Delta_M)$ is the prior on the hyperparameters;
$\likelihood (\{d^{\detone}\}_i | \theta_{i}, \Delta_M)$ is the likelihood of
observing the data $\{d^{\detone}\}_i$ given the source parameters $\theta_{i}$
and the hyperparameters $\Delta_M$;
and $\prior (\theta_{i})$ is the prior on the source parameters for event $i$.
Refer to Appendix~\ref{app:Hierarchical Bayesian Model} for the description of the
hierarchical Bayesian model.

Since the templates can depend on the source parameters, template choice
also determines how $\alpha$ appears in events with different source parameters. 
The important question is whether one can take into account this dependence in the recovery
of $\Delta_{M}$.
We explore this in Appendix~\ref{app:Template source parameter dependence}.
The crucial point is that the BM SNR $\alpha$ will capture the dependence on the source parameters
$\theta$ regardless of the choice of $Z(t; \theta)$, given two conditions: (i) the BM signatures 
remain small (so only leading order effects are relevant) and (ii) the dependence on the template on the source parameters does not change significantly, which can obscure any deviation due to the BM signature present.


\section{Example application of \score}%
\label{sec:Toy model}

In this section, we introduce a model to study the \score framework described in 
the previous section.
The goal is to show that a commonly studied type of BM signature can be recovered using the \score method.
We define hyperparameters $\Delta_M$ which regulate the dependence of the BM signature on chirp mass $\Mc$
as a decaying power law with index $D$.
We model the BM signature $\delta$ induced by this mass-dependent BM signature as a
phase shift in the waveform of a binary black hole (BBH) system parameterized by
the parameter $\beta$ given by
 \begin{align}
	 \label{eq:toy_model_scaling}
	 \beta
	 &=
	 \beta_{0}
	 \left(
		 \frac{\Mc}{M_{0}}
	 \right)^{-D},\\
	 \Mc
	 :&=
	 \left( m_{1} m_{2} \right)^{3/5} \left( m_{1} + m_{2} \right)^{-1/5},
\end{align}
where $\beta_{0}$ is the value of $\beta$ at the reference mass scale $M_{0}$
(we set $M_{0}=5$ M$_\odot$ in the rest of this work), and $m_1$ and $m_2$ are the component masses.
One could also, for example, consider a change in the amplitude of the waveform, but
we focused on the phase shift as it is observed in certain alternatives to GR 
(see full non-linear solutions obtained in BBH mergers within
Einstein-Scalar-Gauss-Bonnet and quartic EFT gravity
theories~\cite{Corman:2022xqg,Cayuso:2023xbc}). Indeed certain pipelines, such
as the
TIGER~\cite{liGenericTestStrong2012,agathosTIGERDataAnalysis2014a,meidamParametrizedTestsStrongfield2018b}
and FTI~\cite{mehtaTestsGeneralRelativity2023} frameworks, search for
deviations in the phase of the waveform when compared to
GR~\cite{sangerTestsGeneralRelativity2024}.

The injected BM signature is then given by
\begin{align}
	\label{eq:toy_model_time_domain}
	\delta \left( t \right) 
	&=
	h_\text{GR} \left( e^{i \phi _\text{T} (t)} - 1 \right) ,\\
	\label{eq:signal_model}
	\phi _\text{T}  
	&= 
	- \beta f(t),\\
	f(t)
	&=
	\frac{16}{3}\beta
	c_{\rm Newt} 
	\frac{M^{2}}{m_{1} m_{2}}
	x^{5/2} \ \times \\\nonumber
	& \frac{1  + n_1 x + n_{3/2} x^{3/2} + n_2 x^{2} + n_{5/2} x^{5/2}+n_3 x^3}
	      {1+ d_1 x + d_{3/2} x^{3/2}},\\
	x :&= \left( \hat{\omega}/2 \right)^{\frac{2}{3}},
	\quad
	\hat{\omega} = M \partial_{t} \phi \left( t \right)
\end{align}
where $\phi$ is the phase of the GR (model) waveform and $M$ is the total mass
of the system. We have used the time dependence of the phase described
in~\cite{dietrichMatterImprintsWaveform2019,dietrichImprovingNRTidalModel2019,colleoniIMRPhenomXP_NRTidalv2ImprovedFrequencydomain2023},
with the parameter $\beta$ acting as the effective tidal deformability
($\tilde{\Lambda}$).
The waveform model $h_\text{GR}$ is the signal predicted by GR. This is
the BM waveform $m = h_\text{GR} $ in the formalism of \score.
The constants $c_{\rm Newt}$, $n_i$, and $d_i$ are fitting values
given in the above references.
This is motivated by the observation that black holes in Effective Field Theories
(EFTs) of gravity described by Eq.~\ref{eq:eft_model} (which we describe below)
have non-zero tidal Love numbers (e.g.~\cite{cardosoBlackHolesEffective2018}), 
but in more general terms describes a phase shift that becomes more consequent
near merger.

A critical feature of the \score framework is that, although the parameter $D$ informs
on the physics of the BM signature, its value alone does not uniquely determine the
source of the BM signature.
Several different physical effects could lead to the same value of $D$. 
As hinted above, one possible source of BM signature that would give a scaling of this
form  is the presence of EFT of gravity effects.
A BM signature due to a higher curvature term in the action of gravity is
expected to scale inversely with the mass of the binary. 
Consider the Lagrangian for the EFT of gravity
\begin{align}
	\label{eq:eft_model}
	{\cal L}_{\rm EFT}
	&=
	\frac{1}{16 \pi G}
	\int 
	d^{4} x
	\sqrt{-g}
	\left( 
		R  
		+
		\cut^{2} \mathcal{L}_{\rm 4}
		+
		\cut^{4} \mathcal{L}_{\rm 6}
		+
		\cdots
	\right),
\end{align}
where $\cut$ is the cutoff length scale of the EFT, and
${\cal L}_{\rm n}$ are the mass dimension-$n$ operators composed of 
functional of the Riemann tensor.
This framework is described in~\cite{endlichEffectiveFormalismTesting2017}.
The physical effect of each group of operators is suppressed by a factor of 
$\left(\cut / M \right)^{D}$, where $M$ is the mass scale of the system and $D$
depends on $n$ in some way that can be computed from the form of the operator.
Current constraints on the scale $\cut$ (for Einstein-scalar-Gauss-Bonnet
theory,
see~\cite{lyuConstraintsEinsteindilationGaussBonnetGravity2022,perinsImprovedGravitationalwaveConstraints2021})
indicate that it must be shorter than the curvature scale of the smallest black
holes observed. 

The idea that BM effects scale as the mass of the system has previously been used in null tests using
post-Newtonian deviation coefficients~\cite{payneCurvatureDependenceGravitationalwave2024} and ringdown 
constraints~\cite{carulloEnhancingModifiedGravity2021a,maselliBlackHoleSpectroscopy2024}.
We stress that this is not the only possible dependence. Another example is an
unaccounted object deformability, which changes the quadrupole moment of the
system with a degree that increases with the spin of the object. This is
notably interesting to distinguish black holes from exotic compact objects
(e.g.~\cite{ryanGravitationalWavesInspiral1995,poissonGravitationalWavesInspiraling1998,laarakkersQuadrupoleMomentsRotating1999,theligoscientificcollaborationTestsGeneralRelativity2021a,krishnenduTestingBinaryBlack2017,Lyu:2023zxv}).

We assume the GW source parameters denoted by $\theta$ (in this analysis we consider the primary and secondary masses $m_1$ 
and $m_2$, the effective spin $\chieff$, and the redshift $z$) are distributed according to the phenomenological
models found to be consistent with the data
in~\cite{abbottBinaryBlackHole2019,abbottPopulationPropertiesCompact2021,abbottPopulationMergingCompact2023}.
The most consequent parameters for the analysis are the component masses, the
redshift, and the spin of the source. The primary and secondary masses, $m_1$
and $m_2$, are drawn from the Power Law + Gaussian peak model introduced 
in~\cite{talbotMeasuringBinaryBlack2018}, which models the observed paucity of
black holes at low masses, a peak at a mass set by the Pair-Instability
Supernova scale, and a power law tail at high masses.
The redshift $z$ is drawn from a distribution inspired by 
the Madau-Dickinson star formation rate, a phenomenological fit of the star
formation rate density of the universe~\cite{madauCosmicStarFormationHistory2014},
as in~\cite{fishbachDoesBlackHole2018}. We have used Flat LCDM cosmological model in this analysis \cite{2011ApJS..192...18K, Planck:2018vyg}. 
The effective spin $\chieff$ is drawn from a Gaussian distribution with mean
$\mu_{\chi}$ and standard deviation $\sigma_{\chi}$. These models are
summarized in Table~\ref{tab:population_parameters} and described in
Appendix~\ref{sec:population_models}.

\begin{table*}
    \begin{center}
        \begin{tabular}{ | c | c | c | }
            \hline
            Physical quantity & Model distribution & Model parameters \\        
            \hline \hline
            Redshift Evolution & 
            \begin{tabular}[t]{@{}c@{}} 
            Madau-Dickinson star formation rate and\\ Flat LCDM cosmological model
             \end{tabular} 
            &
            \begin{tabular}[t]{@{}c@{}} 
                         $\gamma = 2.7$, $\kappa = 2.9$,  $z_{p}=1.9$, \\
			 $\Omega_{M}=0.3$, $\Omega_{\Lambda}=0.7$, $H_{0}=70$ km/s/Mpc
            \end{tabular}  
	    \\
	    \hline
            Primary mass & Power law + Gaussian peak &
            \begin{tabular}[t]{@{}c@{}} 
            $\lambda=0.04$, $\alpha=3.4$, $m_\text{min} =5M_{\odot}$, $m_\text{max} = 100M_{\odot}$, \\  
            $\mu_{m}=35M_{\odot}$, $\sigma_{m}=3.9M_{\odot}$, $\delta_{m}=4.8$  
            \end{tabular}  
            \\  
	\hline
            Secondary mass & Truncated power law & $\beta=1.3$\\
	    \hline
            Effective spin & Gaussian & $\mu_{\chi}=0.06$, $\sigma_{\chi} = 0.12$ \\
             \hline
        \end{tabular}
    \end{center}
    \caption{The astrophysical population parameters and cosmological parameters used in the analysis are given in the table.}
    \label{tab:population_parameters}
\end{table*}

\section{Results}%
\label{sec:Results}

\subsection{CRPS scaling}%
\label{sub:CRPS scaling}

As described in Sec.~\ref{sub:Modelling dependence on source parameters}, 
the SNR $\alpha$ carries a dependence on the source parameters in the presence 
of a BM signature.
Notice three effects that determine the dependence of $\alpha$ on the chirp mass
in the toy model from Sec.~\ref{sec:Toy model}.
First, $\alpha$ depends on $\Mc$ through  $\beta$, and that dependence is
$\Mc^{-2D}$ to leading order in $\beta$.
Secondly, $\alpha$  also depends on the $\Mc$ through the SM model
$h_\text{GR}$, which gives an additional dependence as $\Mc^{5/2}$.
This means that the expected scaling of $\alpha$ with $\Mc$ is
$\Mc^{5/2 - 2D}$.
Finally, the other source parameters on which $\alpha$ does not explicitly 
depend will also affect the BM SNR. As the source
parameters of the GW events arise from an astrophysical population of
parameters described in Table~\ref{tab:population_parameters} which control the
SNR of the event, we expect a spread in the value of $\alpha$ for a given
$\Mc$.

\begin{figure*}
	\begin{minipage}{0.49\textwidth}
		\includegraphics{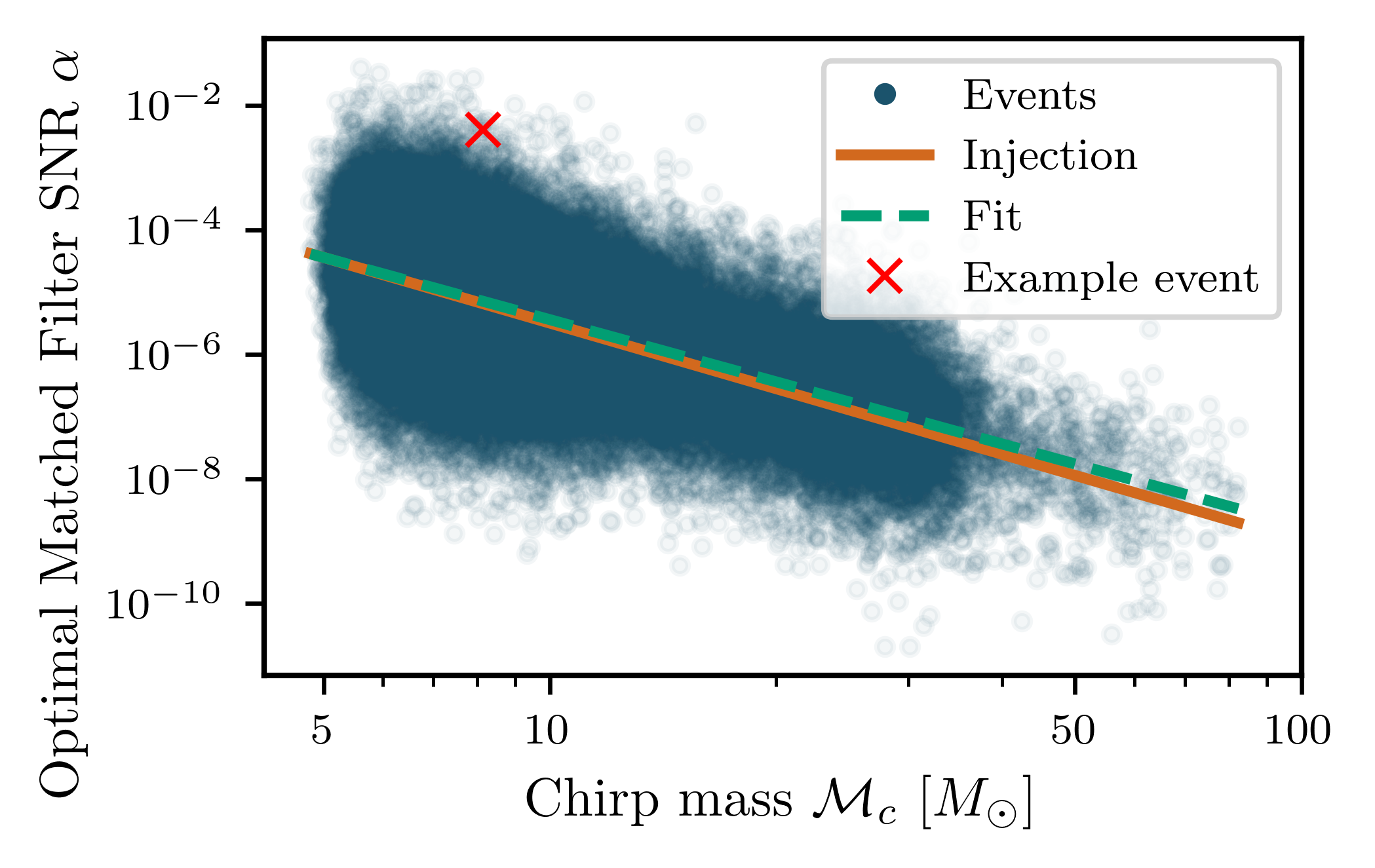}
	\end{minipage}
	\begin{minipage}{0.49\textwidth}
		\includegraphics{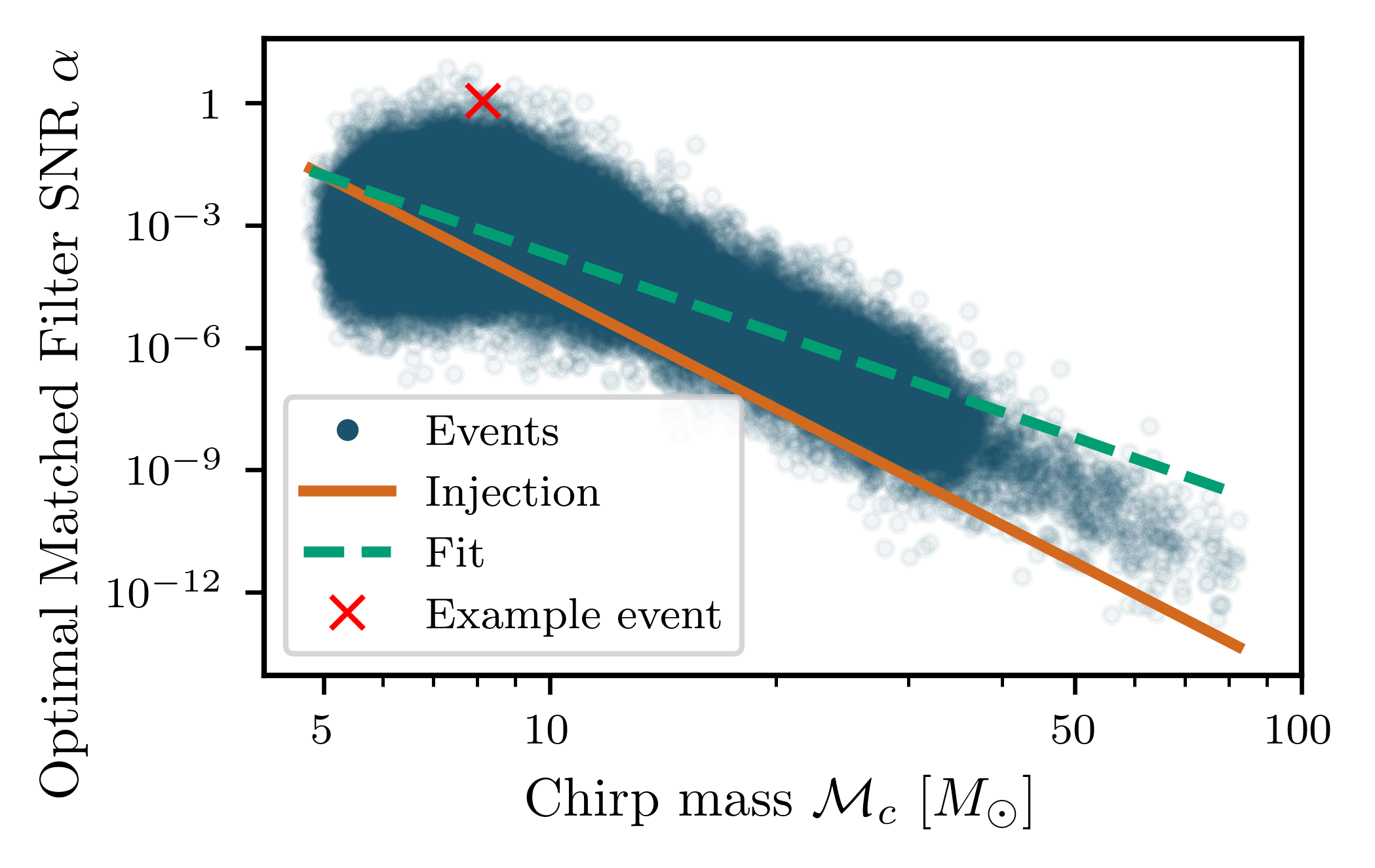}
	\end{minipage}
	\caption{Optimal BM SNR $\alpha$ as a function of the
		chirp mass for an ensemble of $10^{5}$ events. The orange line
		represents the expected scaling of $\alpha$ with $\Mc$. The SNR
		at reference chirp mass $M_{0}$ has been fit (least squares).
		The dashed green line shows the recovered fit. The distribution
		in source parameters other than mass causes a Log-Normal spread
		in the SNR. On the left (right) of the figure, the injected
		scaling is $D=3(6)$ and $\beta_{0}=5(660)$, while the
		least-squares fit recovers $D=2.9(4.5)$. The larger discrepancy
		in the second case is due to the presence of stronger BM
		signatures at lower masses as higher orders of $\beta$ become
		relevant. 
		The BM and SM waveforms, as well as the CRPS for the event with the SNR
		indicated by the red cross, are shown in Fig.~\ref{fig:waveforms}.
	}
	\label{fig:scaling}
\end{figure*}

We illustrate these effects in Fig.~\ref{fig:scaling}, where we plotted the
optimal BM SNR $\alpha$ as a function of the chirp mass for an
ensemble of $10^{5}$ events. The orange line represents the expected scaling of
$\alpha$ with $\Mc$. The SNR at reference chirp mass $M_{0}$ has been fit
(least squares). The dashed green line shows the recovered fit. The
distribution in source parameters other than mass causes a Log-Normal spread in
the SNR. This is because the source parameters affect the waveform through
terms that are multiplied by one another. By the central limit theorem for
products of random variables, the distribution of the product of the terms will
be Log-Normal. 

On the left (right) of Fig~\ref{fig:scaling}, the injected scaling
is $D=3(D=6)$, while the least-squares fit recovers $D=2.9(D=4.5)$. The larger
discrepancy in the second case is due to the presence of
stronger BM signatures at lower masses and higher orders of $\beta$ becoming relevant
\footnote{This is not related to the different values of $D$, but purely to the
strength of the deviation in individual events.}.
The discrepancy between the injected and recovered scaling, when the signal is large,
is due to the presence of next-to-leading order terms in $\beta$ contributing to 
the BM SNR Eq. ~\eqref{eq:def_power_projection}. The value of $\beta$ at which
higher than leading order terms become relevant depends on the behavior of $f(t)$. 

In the low $\beta$ case where only the leading-order effect is important,
the BM SNR scales in a way that can be predicted directly from expecting a power-law,
and therefore one can recover the $D$ without fine-tuning the template $Z$.
In the case of high $\beta$,  next-to-leading order terms must be included in the template
to improve the recovery of the scaling, as described in the \texttt{SCoRe} technique \cite{dideronNewFrameworkStudy2023}. The latter case is less agnostic to the choice of template.

\begin{figure*}
	\begin{minipage}{0.49\textwidth}
		\includegraphics{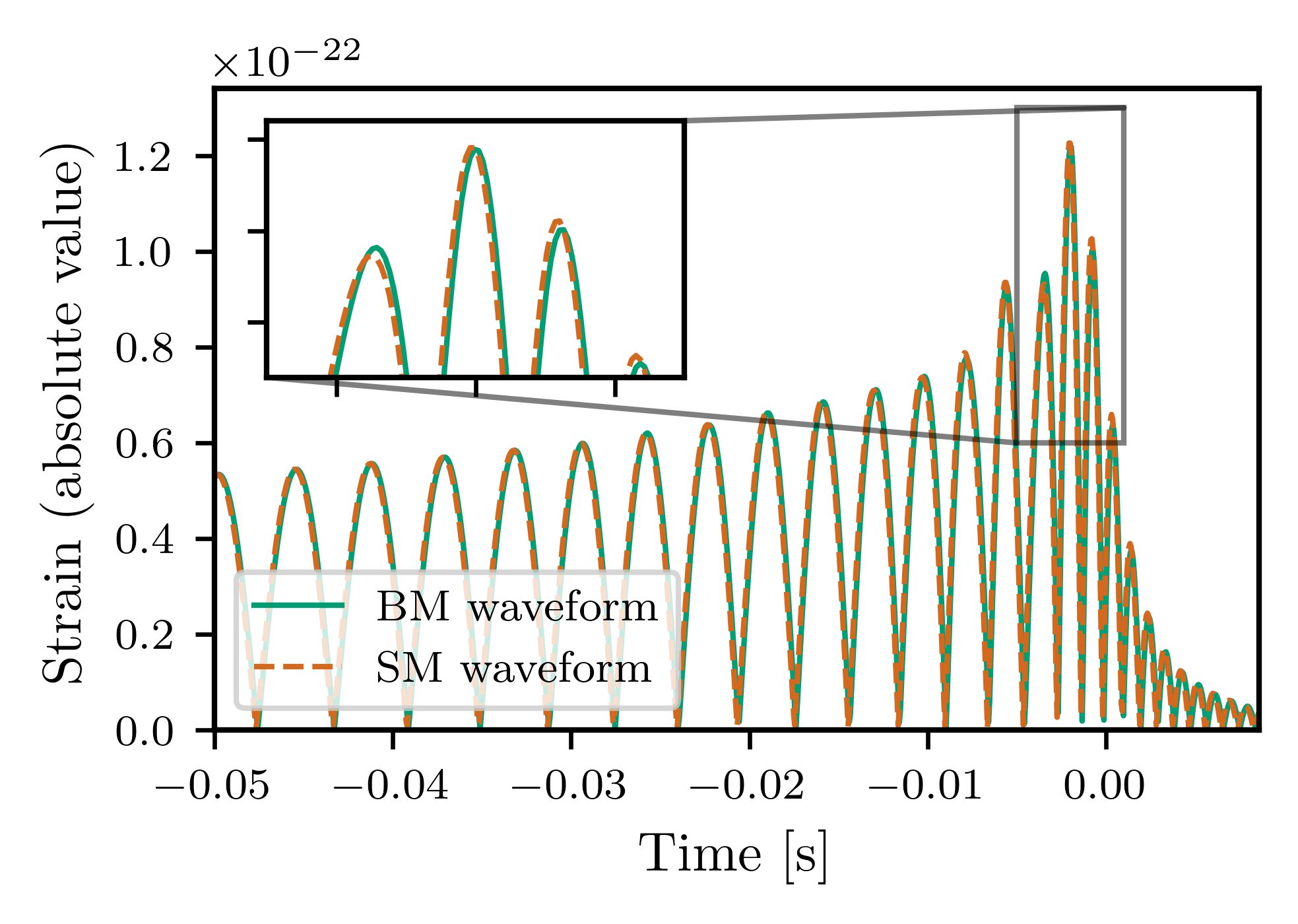}
	\end{minipage}
	\begin{minipage}{0.49\textwidth}
		\includegraphics{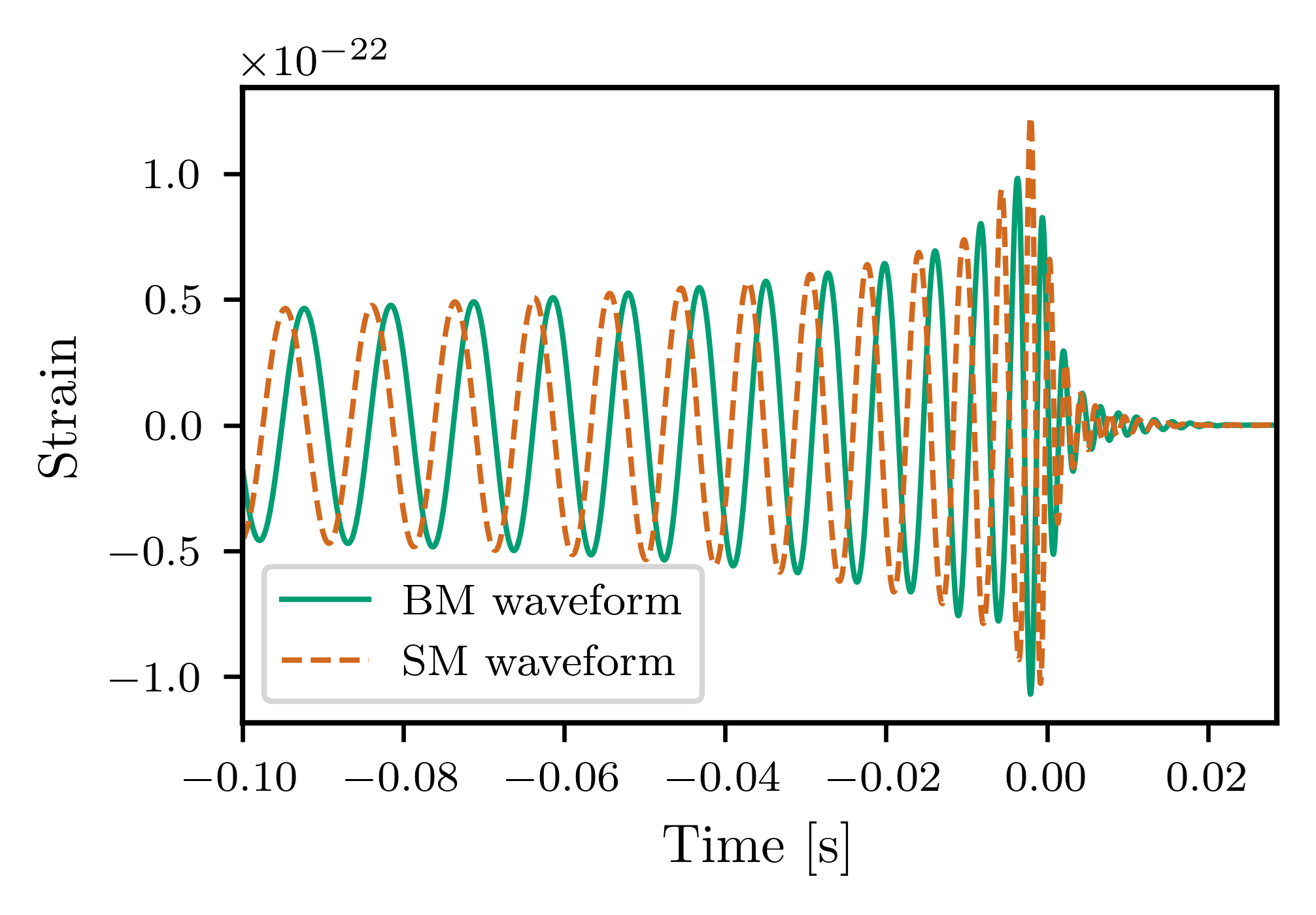}
	\end{minipage}
	\begin{minipage}{0.49\textwidth}
		\includegraphics{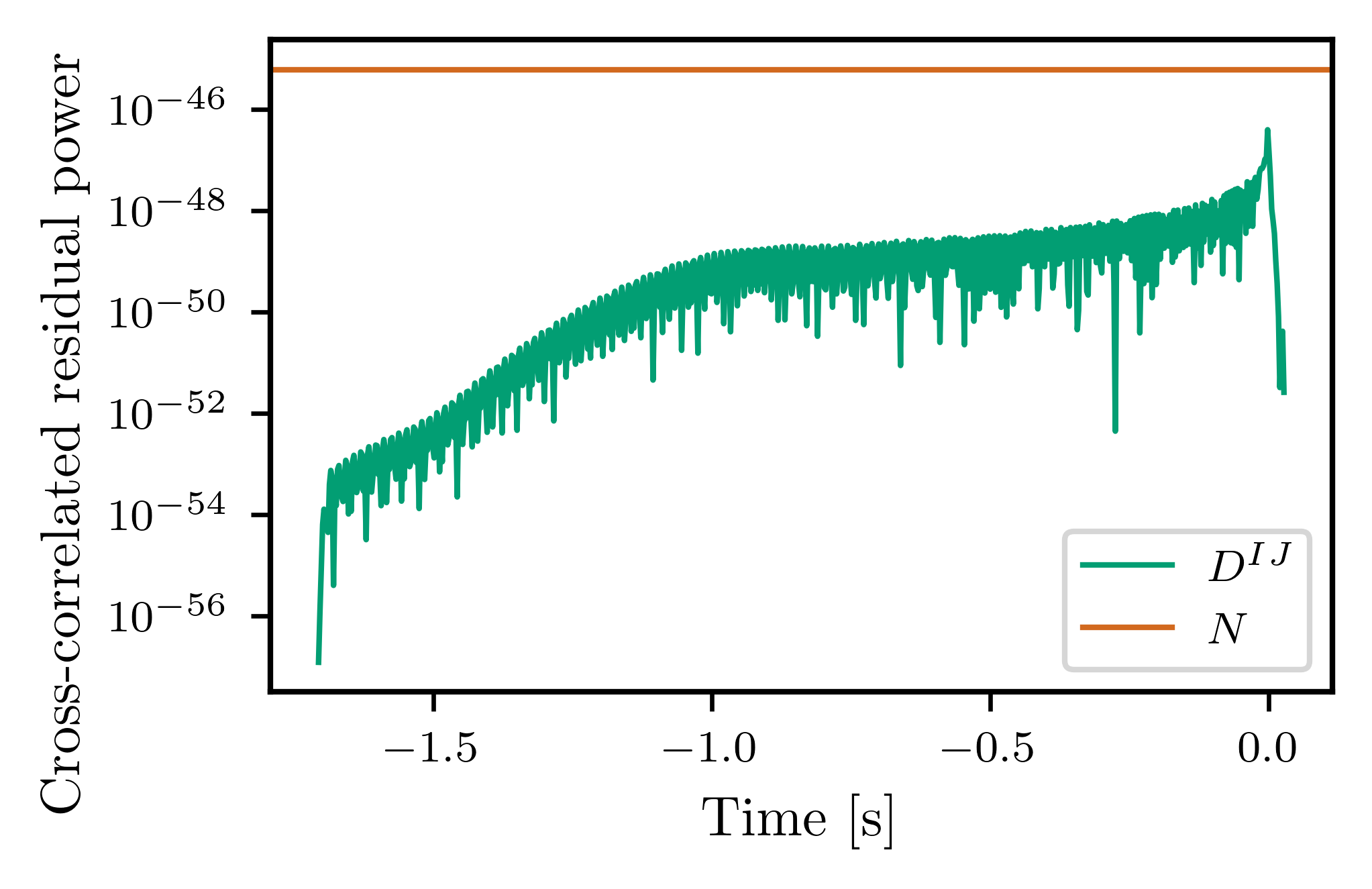}
	\end{minipage}
	\begin{minipage}{0.49\textwidth}
		\includegraphics{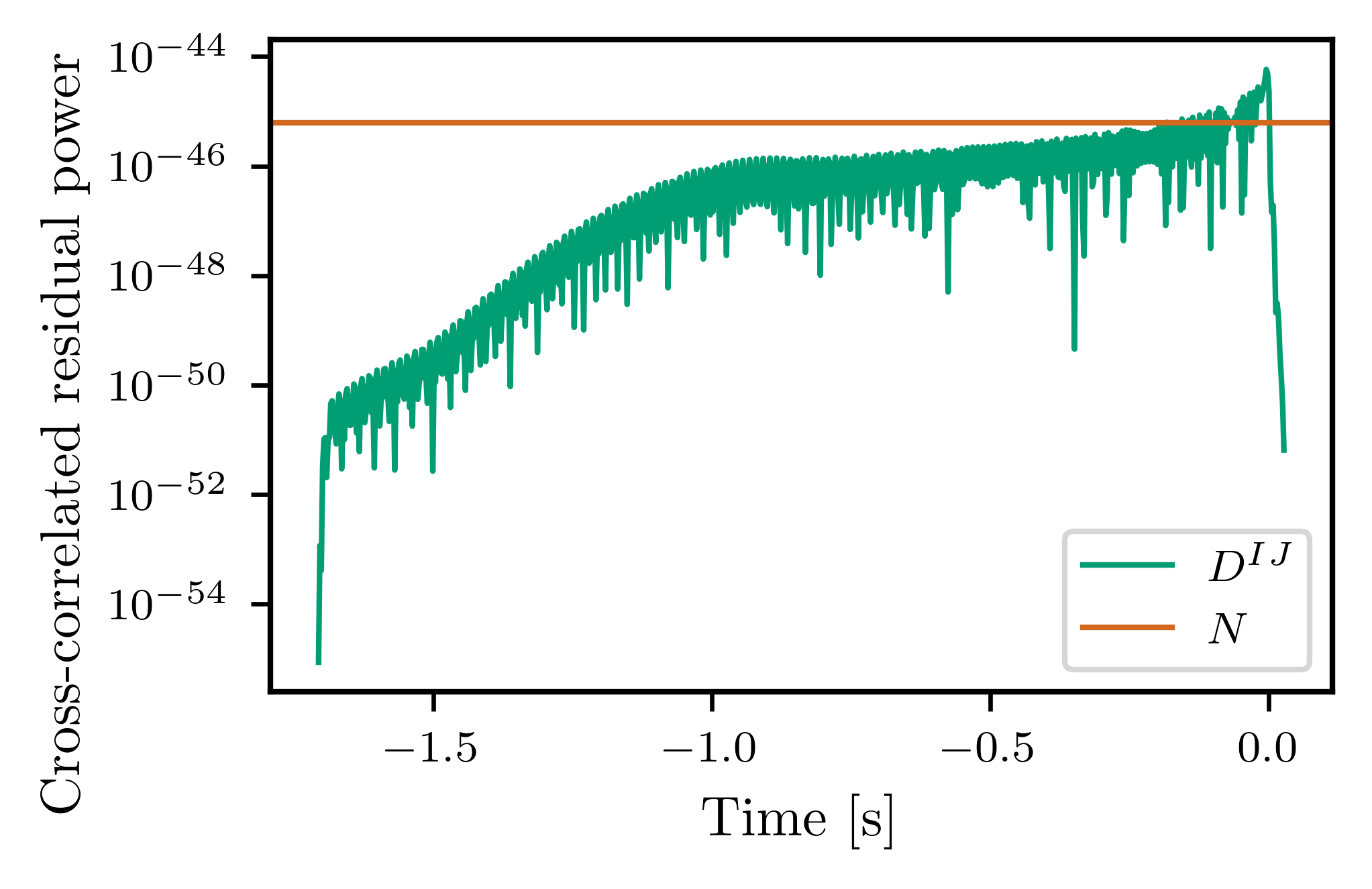}
	\end{minipage}
	\caption{Top row: SM and BM waveforms for the example event indicated by a red cross in Fig.~\ref{fig:scaling}.
		The event has a luminosity distance of \exampleEventdL and chirp mass of \exampleEventMc.
		The strain shown has been projected onto the illustrative New Mexico CE detector.
		The case on the right corresponds to the steeper power law ($D=6$), and gives $\beta=\exampleEventLargeBeta$ for this event, while on the left $D=3$ gives $\beta=\exampleEventSmallBeta$. 
		Bottom row: CRPS of the event, plotted for both values of $\beta$.
		For the larger value of $\beta$, the CRPS surpasses the cross-correlated noise level near the merger.
	}
	\label{fig:waveforms}
\end{figure*}

The difference between the high and low $\beta$ cases 
is illustrated in more detail in Fig.~\ref{fig:waveforms}, 
where, in the first row, we illustrate the SM and BM waveforms
for the example event indicated by a red cross in Fig.~\ref{fig:scaling}.
The event had a luminosity distance of \exampleEventdL\, and chirp mass
of \exampleEventMc, and the strain shown was projected onto 
the illustrative New Mexico CE detector.
The case on the right corresponds to the steeper power-law $D=6$, and gives
$\beta=\exampleEventLargeBeta$ for this event, while on the left $D=3$ gives
$\beta=\exampleEventSmallBeta$. 
The stronger value of $\beta$ gives a visually recognizable phase shift in the 
waveform. The CRPS of the event is plotted for both values of $\beta$ on the 
second row of Fig.~\ref{fig:waveforms}.
For the larger value of $\beta$, the CRPS surpasses the cross-correlated noise
level near the merger.
To compute the CRPS in Figs.~\ref{fig:scaling} and~\ref{fig:waveforms}, we have
used $\tau=2.44$ ms, corresponding to integrating over 40 time-samples at 16kHz.

\subsection{Constraints for next-generation network of GW detectors}%
\label{sub:Constraints for CE-like network}

The last section argued that BM hyperparameters can be recovered from a population
of events, given that the BM signature is small in individual events.
We now consider how well our example BM model hyperparameter $D$ can be constrained
by a network of GW detectors.
We are interested in demonstrating the method in the regime where the SNR of
the SM model is high, but much larger than the SNR of the BM signature---where
the residual signal is small compared to the SM model. This regime will be
relevant for the next-generation detectors, so we consider a network of two
CE-like detectors. The same analysis could be repeated for current detectors,
especially by using data from already detected events.

We use the Power Spectral Density (PSD) derived from the 
Amplitude Spectral Density (ASD) for the baseline 40 km CE design~\cite{srivastavaSciencedrivenTunableDesign2022a}
and the New South Wales and Idaho locations in~\cite{borhanianGWBENCHNovelFisher2021a}.
We have used the CE PSD and potential configurations as an illustration
of 3G detector capacities, but the analysis could also be performed with an ET-like detector
(see for example~\cite{hildSensitivityStudiesThirdgeneration2011,branchesiScienceEinsteinTelescope2023a}).
As the anticipated sensitivity of both types of detectors is similar, we do not expect the 
results to differ qualitatively.

In this analysis, the source parameters of GW events are drawn from an astrophysical population model (as described in the Appendix \ref{sec:population_models}) with a choice of parameters shown in
Table~\ref{tab:population_parameters}.
Events with optimal match-filtering SNR (computed using the SM waveform $h_\text{GR}$)
greater than 8 are selected. 
The likelihood $\likelihood$ in Eq.~\ref{eq:hierarchical_posterior} is then computed 
by marginalizing over the source parameters and the BM SNR $\alpha$.
As we are interested in the precision of the method, we do not compute the full likelihood for each
event.
Instead, we approximate the likelihood in $\alpha$ and  $\theta$ as a Gaussian centred
on the true values---the values drawn from the population models for $\theta$
and at zero for $\alpha$.
The variance of the Gaussian is computed using the Fisher matrix for the best template
(the template Z(t) computed using the injected signal) and the SM waveform model 
for the event. 
This yields a lower limit on the variance.
See Appendix~\ref{app:Fisher analysis} for the details of this analysis.

\begin{figure}[htpb]
	\centering
	\includegraphics[width=\columnwidth]{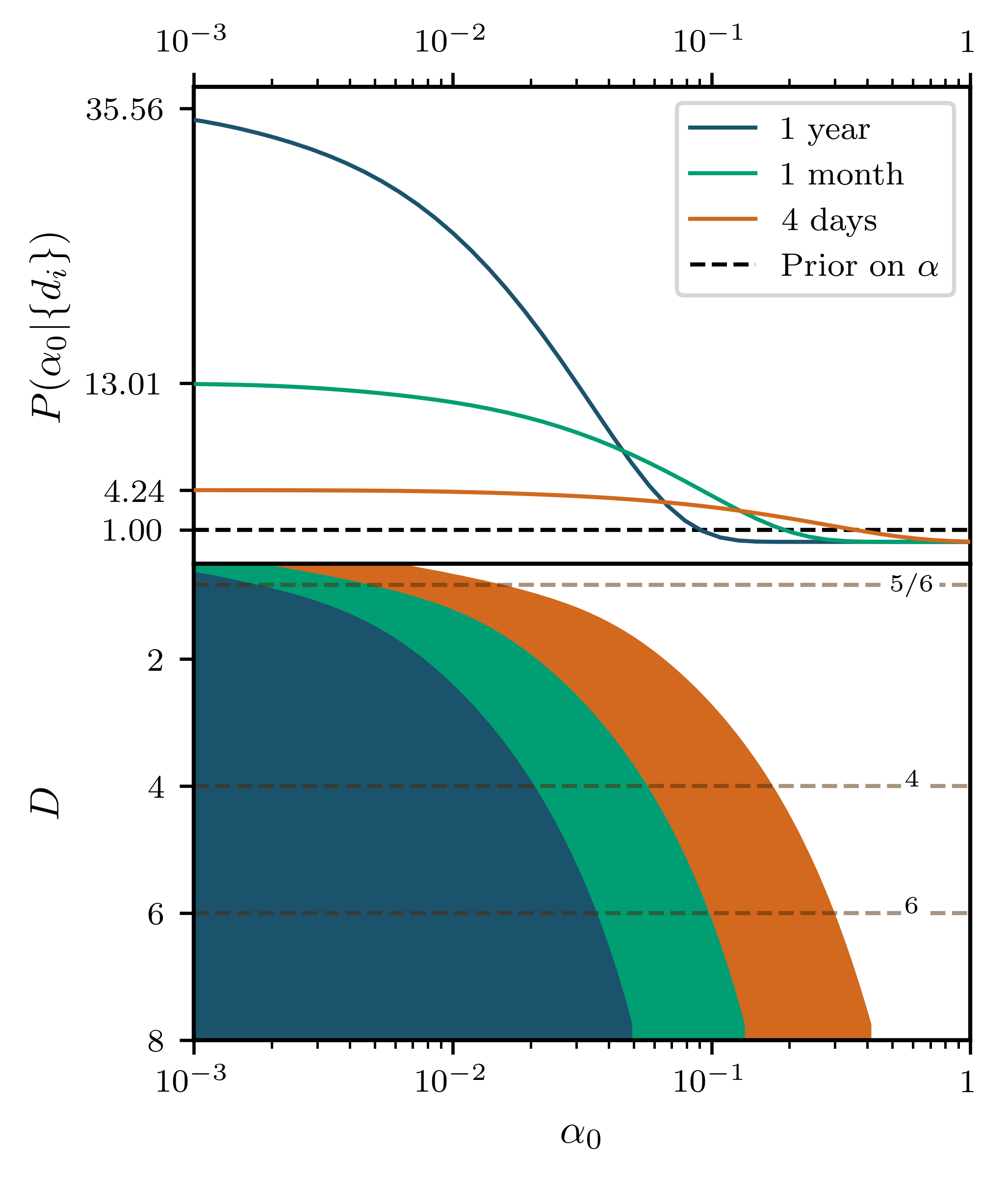}
	\caption{Posterior on $D$ and $\alpha_{0}$ for a network of two CE-like detectors.
		The lower plot shows the one $\sigma$ credible region of the joint
		posterior for different observation times (assuming a detection rate of
		$10^{5}$ per year).
		The dashed lines mark the values of the illustrative model for BM deviations.
		For example, certain scalar-tensor theories give $D=4$, while
		vacuum solutions in Quartic Gravity give $D=6$. 
		Unaccounted eccentricity effects would give $D=5/6$. The constraints
		The constraints loosen as $D$ increases and tighten as the square root of observation time (the number of events).
		The upper plot
		shows the posterior marginalized over $D$. In the absence of 
		BM signatures, $\alpha_{0}=0$, and the prior on $\alpha_{0}$ is flat from $0$ to $1$.
		For these two reasons, the value of the marginalized posterior at zero is a measure
		of the evidence for the SM as opposed to any BM signature in the CRPS of the combined data
		(the Savage Dickey ratio).
	}
	\label{fig:null_results}
\end{figure}

In Fig.~\ref{fig:null_results}, we show the posterior on $D$ and $\alpha_{0}$, the value 
of $\alpha$ at the reference chirp mass $5M_{\odot}$. The latter is a measure of 
the presence of any BM signature in the data. Flat priors were assumed for both
parameters, ranging from $2$ to $8$ for $D$ and $0$ to $1$ for $\alpha_{0}$.
The left plot shows the upper limit of the $68\%$ credible interval region of the joint
posterior. 
Two features are apparent. First, the constraints loosen as $D$
increases (generally, for higher mass dimension operators). A sharper power law
decay is harder to distinguish from the SM model as the BM signature is weaker
for most masses.
The constraints tighten as the square root of the number of events. We have
assumed a detection rate of $10^{5}$ per year~\cite{regimbauDiggingDeeperObserving2017}.
The right plot shows the posterior marginalized over $D$. 
As $\alpha_0$ is a measure of the presence of any BM signature in the data (and
its prior is flat from $0$ to $1$), the intercept of its posterior with zero
gives the Savage Dickey ratio, which is a measure of the evidence for the SM
as opposed to any BM signature in the CRPS of the combined data.
The zero-intercepts for the different observation times are shown on the
x-axis.

We mark the values of the illustrative model for BM deviations with the dashed lines.
For example, certain scalar-tensor theories, such as Einstein scalar
Gauss-Bonnet gravity predicts $D=4$, while vacuum solutions in Quartic
Gravity predict $D=6$. Known physics could also give rise to BM
signatures scaling if not accounted for in waveform models. For example,
eccentricity effects would give
$D=5/6$~\cite{sainiSystematicBiasParametrized2022,sainiEccentricityinducedSystematicError2024}.
Although this was not included in our prior, we note that this effect could potentially be better constrained
than the integer values of $D$ corresponding to EFT operators.
We note that, although scaling may be recovered or constrained using this method, it is
not possible to distinguish between different models giving the same scaling.

\subsection{Mass range influence on inference}%
\label{sub:influence}

The results presented in the previous section combine the information from events of
all source parameters, particularly masses. In the specific example for which we demonstrated the method, the strength of the BM
signature decreases with increasing source mass. At the same time, due to the Power-Law + Gaussian mass
distribution and GW selection effects \footnote{The fraction of lighter mass sources over the high mass sources detectable up to high redshift (and hence over a larger cosmic volume) is less.}, only a few lighter masses are detectable. As a result, the strongest
BM signals may be evident in only a few sources and thus not contribute significantly to the constraints.

To illustrate these two effects and how they impact the influence different mass ranges 
have on the inference of $D$, we consider the posterior on $D$ obtained
from considering only events with a chirp mass within a window of width $\delta \Mc=\influenceDeltaM$.
To focus on the effect of the mass distribution and the variation of SNR with mass, 
we make three assumptions for the purpose of efficiently computing the posteriors Eq. ~\eqref{eq:hierarchical_posterior}, (i)
we assume that the SM source parameters are known exactly, (ii) that the SNR of the BM signature depends only on the chirp mass, and (iii) we are in the limit of a large number of events (as expected from the next generation detectors). The computation of the posterior using these assumptions
is described in Appendix~\ref{app:Mass range influence on inference}. For the BM signal, we inject an optimal BM SNR $\alpha$ given by a power law in the
chirp mass $\alpha = \alpha_{0} \left( \Mc / 5M_{\odot}
\right)^{\frac{5}{2}-2D}$, where $\alpha_{0}$ is known exactly and set to \influenceAlpha
\footnote{The value of $\alpha_0$ for an actual BM signature would be set by the template used
to filter the CRPS and the noise properties of the detector network.}
.

\begin{figure}
	\centering
	\includegraphics[width=\columnwidth]{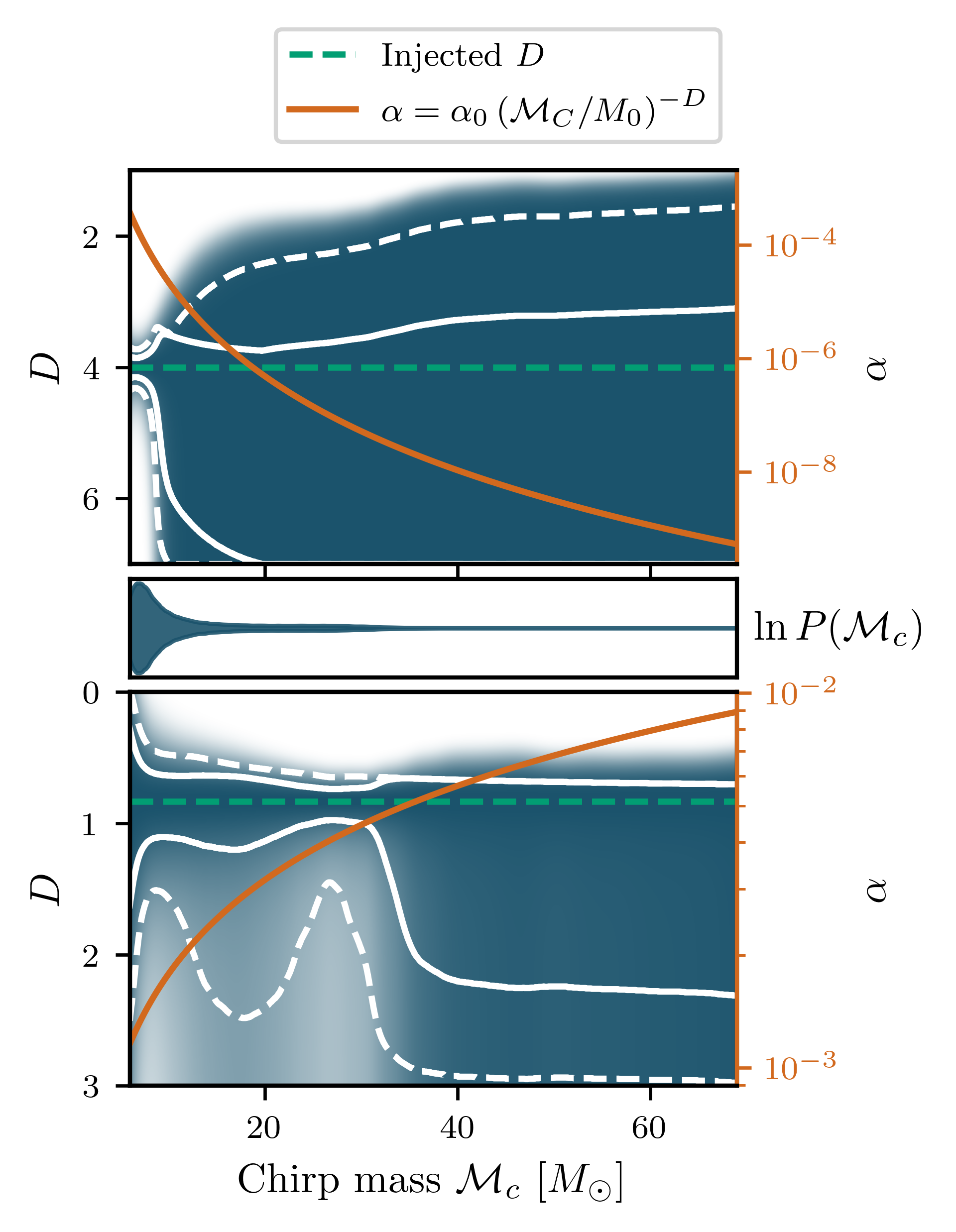}
	\caption{
		Posterior on $D$ for different mass ranges (we assume the value of $\alpha_{0}$
		is known exactly in the inference).
		. The blue intensity shows the value of the posterior
		$\posterior (D | \{d^{\detone}\}_i)$ computed for events with chirp mass within $[\Mc - \delta \Mc, \Mc + \delta
		\Mc]$ (each value of the x-axis has its own
		posterior---in the same fashion as a violin plot).
		The white lines show the upper and lower limits of the 1 (solid) and 2 (dashed)
		$\sigma$ credible intervals change for different mass ranges.
		The shorter the credible interval, the tighter the constraints on $D$.
		We overlay the injected SNR as the orange line
		corresponding to the right axis. The mass distribution is shown
		in the middle plot. In the upper (lower) plot,
		we injected $D=\influenceDHigh\left(\influenceDLow\right)$. The total
		number of events in each plot is set so that the tightest
		constraints are visually comparable (it is set to
		\influenceNEventsHigh(\influenceNEventsLow) for the upper (lower) plot). 
		For a steeply decaying power law (with $D=\influenceDHigh$), low mass ranges
		are significantly more informative than higher masses. For shallower or 
		increasing power laws (with $D>\frac{5}{4}$), the mass distribution
		determines the influence of different mass ranges.
	}
	\label{fig:influence}
\end{figure}

In Fig.~\ref{fig:influence} 
, the blue intensity shows the value of the posterior $\posterior (D | \{d^{\detone}\}_i)$
computed for events with chirp mass within $[\Mc - \delta \Mc, \Mc + \delta
\Mc]$ (this is not a joint posterior, as each value of the x-axis has its own
posterior---in the same fashion as a violin plot).
The white lines show the upper and lower limits of the $68\%$ (solid) and $95\%$ (dashed)
credible intervals change for different mass ranges.
The shorter the credible interval, the tighter the constraints on $D$.
To display the two factors affecting the inference, we overlay the injected
the injected power law as the orange line corresponding to the right axis. The
mass distribution is shown in the middle plot.
In the upper (lower) plot, the injected $D=\influenceDHigh\left(\influenceDLow\right)$. 
The total number of events in each plot is set so that the tightest constraints 
are visually comparable (it is set to
\influenceNEventsHigh(\influenceNEventsLow) for the upper (lower) plot). Changing
the total number of events would move the constraints in the same proportion
across all mass ranges.

The salient point of Fig.~\ref{fig:influence} is that different mass ranges
become more or less informative depending on the scaling $D$.
In the case of a steeply decaying power law (with $D=\influenceDHigh$), low
masses are significantly more informative than higher masses regardless of
relative abundance. In the upper plot, the constraints on $D$ are shallowest 
at the lowest chirp masses and then increase with mass. Note the slight 
convexity of the upper constraints around the PISN peak of the mass distribution 
($\Mc \approx 35M_{\odot}$). 
As the power law becomes shallower or even increasing (for any $D>5 / 4$), 
the influence of different mass ranges becomes more dominated by the mass distribution.
In the lower plot, where we injected $D=\influenceDLow$), the constraints are tightest at the initial
mass peak ($\Mc \approx 10M_{\odot}$) and at the PISN peak. The lowest mass ranges are 
not the most informative.
In both cases, the decaying power law in the mass distribution means that
the constraints on $D$ are weakened at higher masses
\footnote{
	There is an subtle increase in the mass distribution at $\Mc \approx 50M_{\odot}$
	due to asperities in the KDE estimate (see Appendix~\ref{app:Mass range influence on inference}).
}
.

\section{Final words}\label{sec:Conclusions}

Whether due to new physics or simply to the complexity of the emission of GWs,
our waveform models may not fully capture GW signals in the
real data given the sensitivity detectors can reach in the future. This is
particularly true as the SNR and sheer number of detected GW events increase
with upgrades and new detectors. 
At the same time, this increase in available data opens the way for more data-driven
approaches that do not rely on specific models of BM signatures.
A particularly useful feature of BM signatures that can be probed with more data is that
some depend on the GW source properties.

In this work, we explored the combination
of two such data-driven approaches that can capture any GW source-dependent BM
signatures in a model-independent way. Using a novel BM search technique
\texttt{SCoRe} \cite{dideronNewFrameworkStudy2023}, we show that one can make
an inference of any BM signal by taking into account the dependence on the GW
source properties. The cross-correlation of residual which is used in
\texttt{SCoRe} allows to distinguish the signal from noise without any specific
model assumptions and search for source parameter-dependent BM physics or
systematic.

As an illustration, we used a deviation scaling inversely with chirp
mass (motivated by EFT of gravity arguments) appearing as a phase shift in the
waveform (e.g.~\cite{endlichEffectiveFormalismTesting2017,Cano:2023jbk,scalingeft}).
We showed that the index of the power law scaling could be recovered using next-generation GW detectors.
We note that while extracting such scaling would not allow one to distinguish between specific models (e.g.
specific extensions of GR) will constrain sub-families that are consistent.
Further analysis then, with deeper knowledge of specific deviations in particular theories would enable further scrutiny.  In Sec.~\ref{sub:Constraints for CE-like network}, we forecast the
constraints that a network of two CE-like detectors could place on the power-law index with up to one year of observation time. 
Further, since the variation of the deviation can depend on source parameters and hence impact the inference, we explored which parts of the parameter space are most informative to find BM signatures. 
In Sec.~\ref{sub:influence}, we showed that lower masses are most informative when
the underlying deviation steeply decays with chirp mass, while the mass distribution
dominates the inference when the deviation is shallower or increasing with mass.

Although a specific model (a deviation scaling inversely with mass scale) is
used in this paper to demonstrate the ability of the technique \score to
capture a GW source-dependent signature in the data, it can be applied to any
kind of BM signatures ranging from physical effects to unknown waveform
systematic. This method can be easily extended to look for new physical effects
such as spin-induced quadrupole moments, tidal deformation of neutron stars,
and signatures of exotic compact objects. Furthermore, it can be used for
discovering any common systematic between different detectors induced due to
mis-modeling of waveforms\footnote{In order for their effect on parameter
	inference to remain below that of statistical errors, in CE-like
ground-based detectors, the level of systematic errors should be reduced by
1--3 orders of magnitude~\cite{purrerGravitationalWaveformAccuracy2020}.}.  We
plan to explore these in our future work. In summary, this technique \score
makes it possible to discover BM signatures in a data-driven way and its
application on the current network of GW detectors and in the next-generation
detectors will bring new insights into the Universe.

\acknowledgements
The authors are thankful to Maximiliano Isi for providing useful comments on the manuscript as a part of the LIGO publication and presentation policy and for useful discussions. 
The authors also thank Katerina Chatziioannou, Zoheyr Doctor, Reed Essick, and Ma SiZheng for useful discussions.
The authors would like to thank the LIGO-Virgo-KAGRA Scientific Collaboration for providing the noise curves. 
This research has made use of data or software obtained from the Gravitational Wave Open Science Center (gw-openscience.org), a service of LIGO Laboratory, the LIGO Scientific Collaboration, the Virgo Collaboration, and KAGRA. LIGO Laboratory and Advanced LIGO are funded by the United States National Science Foundation (NSF) as well as the Science and Technology Facilities Council (STFC) of the United Kingdom, the Max-Planck-Society (MPS), and the State of Niedersachsen/Germany for support of the construction of Advanced LIGO and construction and operation of the GEO600 detector. Additional support for Advanced LIGO was provided by the Australian Research Council. Virgo is funded, through the European Gravitational Observatory (EGO), by the French Centre National de Recherche Scientifique (CNRS), the Italian Istituto Nazionale di Fisica Nucleare (INFN) and the Dutch Nikhef, with contributions by institutions from Belgium, Germany, Greece, Hungary, Ireland, Japan, Monaco, Poland, Portugal, Spain. The construction and operation of KAGRA are funded by Ministry of Education, Culture, Sports, Science and Technology (MEXT), and Japan Society for the Promotion of Science (JSPS), National Research Foundation (NRF) and Ministry of Science and ICT (MSIT) in Korea, Academia Sinica (AS) and the Ministry of Science and Technology (MoST) in Taiwan. This material is based upon work supported by NSF's LIGO Laboratory which is a major facility fully funded by the National Science
Foundation. 
This research was supported in part by a NSERC Discovery Grant and CIFAR (LL),
the Simons Foundation through a Simons Bridge for Postdoctoral Fellowships  (SM) as well as Perimeter Institute for Theoretical Physics. 
Research at Perimeter Institute is supported by the Government of
Canada through the Department of Innovation, Science and Economic Development
Canada and by the Province of Ontario through the Ministry of Research,
Innovation and Science. The work of SM is a part of the $\langle \texttt{data|theory}\rangle$ \texttt{Universe-Lab} which is supported by the TIFR and the Department of Atomic Energy, Government of India.  In the analysis done for this paper, we have used the
following packages: \textsc{PyCBC}~\cite{nitzGwastroPycbcV22022}, \textsc{LALSuite}~\cite{lalsuite}, 
\textsc{NumPy}~\cite{harris2020array}, \textsc{SciPy}~\cite{2020SciPy-NMeth}
and \textsc{Matplotlib}~\cite{Hunter:2007} with
\textsc{Seaborn}~\cite{Waskom2021}.

\bibliography{main.bib}

\appendix	

\section{Hierarchical Bayesian Model}%
\label{app:Hierarchical Bayesian Model}

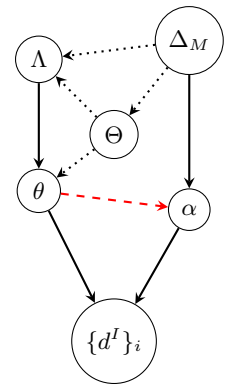
\begin{figure}
    \begin{center}
    \begin{tikzpicture}[node distance=2cm,scale=2]
        \tikzstyle{block} = [circle, text centered, draw=black]
        \tikzstyle{arrow} = [thick, ->, >=stealth]
        \node (lambda) [block] {$\Lambda$};
        \node (theta) [block, below of=lambda, yshift=+0.25cm] {$\theta$};
        \draw [arrow] (lambda) -- (theta);
	\node (beyond gr) [block, right of=lambda, yshift=+0.25cm] {$\Delta_M$};
	\node (beta) [block, below of=beyond gr,yshift=-0.25cm] {$\alpha$};
	\draw [arrow] (beyond gr) -- (beta);
	\node (data) [block, below of=theta,xshift=+1cm] {$\{d^\detone\}_i$};
        \draw [arrow] (theta) -- (data);
	\draw [arrow] (beta) -- (data);
	\draw [arrow,dashed,red] (theta) -- (beta);
	%
	\draw [arrow,dotted] (beyond gr) -- (lambda);
	\node (cosmo) [block, below of=lambda, yshift=+1cm, xshift=+1cm] {$\Theta$};
	\draw [arrow,dotted] (cosmo) -- (lambda);
	\draw [arrow,dotted] (cosmo) -- (theta);
	\draw [arrow,dotted] (beyond gr) -- (cosmo);
    \end{tikzpicture}
    \end{center}
    \caption
    {Schematic representation of the Hierarchical Bayesian model for the
	    inference of a BM signature. 
	    The population hyperparameters $\Lambda$ (e.g PISN mass scale)
	    describe the distribution of source parameters $\theta$ (e.g. mass
	    and spin) in the population. Likewise, the distribution of  BM 
	    signature parametrized by  $\alpha$ measured in individual events
	    is described by the hyperparameters $\Delta_M$. Cosmological
	    parameters $\Theta$ could
	    also affect these distributions. By assuming such a hierarchy 
	    of dependences and marginalizing over the intermediate parameters,
	    we can recover hyperparameters from a set of event data $\{d^\detone\}_i$.
	    An arrow from node $A$ to node $B$ denotes that the prior on $B$
	    depends on $A$.
	    The physics underlying the BM signature is common to all
	    events and determines the deviation $\delta$ in individual events 
	    according to the parameters $\Delta_M$, which should ultimately
	    describe the underlying physics of the BM model (e.g the
	    energy dimension of the leading EFT operator).
	    Although the morphology of the deviation is not known (and must be
	    modeled by the template functions), we expect it to vary
	    consistently with the source parameters $\theta$ (e.g with a power
	    law dependence on the mass of the BH in our
	    EFT-motivated model).
	    We include this dependence, which corresponds to the red-dashed
	    arrow, in our model.
	    We then attempt to infer the parameters $\Delta_M$ from the data
	    $\{d^\detone\}_i$ by marginalizing over the uncertainties in the
	    intermediate parameters $\theta$ and $\alpha$.
	    The source parameters themselves are distributed according to the
	    population hyperparameters $\Lambda$.
	    For the most complete model, we would also include the effect of
	    cosmological models described by parameters $\Theta$, which could
	    depend on the BM model and affect the source population.
    }
    \label{fig:dag}
\end{figure}

In this appendix, we describe the hierarchical Bayesian model used to infer BM
hyperparameters from a set of events, as described in
\cite{dideronNewFrameworkStudy2023}. We expect a BM signature to appear
differently in different events. Assuming the BM signature is parametrized by a
set of parameters $\alpha$, the value
$\alpha$ takes in individual events is determined by a set of hyperparameters
$\Delta_M$ (for example, the scale and power law index described in
Sec.~\ref{sec:Toy model}).
This is shown by an arrow from $\Delta_M$ to $\alpha$ in Fig.~\ref{fig:dag}.
Similarly, the population hyperparameters $\Lambda$ (e.g. the Pair-instability Supernova mass scale) describe the distribution of source parameters $\theta$ (e.g. mass and spin) in the population.
The key assumption in our model is that the way $\alpha$ varies across events is informed by the source parameters $\theta$. 
This is shown by the red dashed arrow in Fig.~\ref{fig:dag}
\footnote{
For completion, we could also include Cosmological parameters $\Theta$ in our model. These may affect both the source population parameters $\theta$ and their distribution (described by $\Lambda$). If the BM signature arises from a modified theory of gravity, the BM hyperparameters may also be related to  $\Theta$. These relations are shown by the dotted arrows in Fig.~\ref{fig:dag}. In a complete model, all these parameters would be included and inferred or marginalized.}
.

This structure can be used to build a hierarchical Bayesian model to infer the hyperparameters $\Delta_M$ from a set of events $\{d\}_i$, each labeled by $i$.
The posterior on the hyperparameters $\Delta_M$ is given by
\begin{align}
	\label{eq:hierarchical_model_1}
	\posterior 
	\left( \Delta_M | \{d\} \right)
	&\propto
	\likelihood 
	\left( \{d\} | \Delta_M \right)
	\prior \left( \Delta_M \right),
\end{align}
where $\{d\}$ denotes the set of data.
Treating the events as independent, the likelihood $\likelihood \left( \{d\} | \Delta_M \right)$ is given by the product of the likelihoods of individual events:
\begin{align}
	\likelihood 
	\left( \{d\} | \Delta_M \right)
	&=
	\prod_{i}^{N_\text{obs} }
	\prod_{\detone \neq \dettwo}^{N_\text{det} }
	\likelihood
	\left( \{d^{\detone}, d^{\dettwo} \}_{i}  | \Delta_M, \mathcal{S}  \right),
\end{align}
where each event labelled $i$ contains a product over pairs of detectors $\detone$ and $\dettwo$.
The product is over the $N_\text{obs}$ observed events, and $\mathcal{S}$, on the left side of the conditioning bar, denotes that an event has been detected.
The selection effects on $\likelihood \left( \{d^{\detone}, d^{\dettwo}\}_i | \Delta_M, \mathcal{S}  \right)$ can be computed taking into account the effects of both $\theta$ and $\alpha$ (see~\cite{dideronNewFrameworkStudy2023}). 
Here, we assume that any event with an optimal match filtering network SNR above \thresholdSNR is detected. 
This SNR is computed using the SM template (therefore only taking into account variation in detectability due to $\theta$).

We then marginalize over the source parameters $\theta$ and the BM signature in individual events ($\alpha$), to obtain the likelihood on the hyperparameters $\Delta_M$:
\begin{align}
	&\likelihood
	\left( 
		\{ d^{\detone}, d^{\dettwo} \}_{i}
		|
		\Delta_{M}
	\right)
	\nonumber
	\\
	=&
	\int
	d \theta_{i} d \alpha_{i}
	\likelihood
	\left( 
		\{ d^{\detone}, d^{\dettwo} \}_{i}
		|
		\alpha_{i}
		,
		\theta_{i}
	\right)
	\prior
	\left( 
	\alpha_{i} | \theta_{i}, \Delta_{M}
	\right)
	\prior (\theta_{i}),
	\label{eq:hierarchical_model_2}
\end{align}
where we have used the fact that the likelihood only depends on $\Delta_{m}$ through $\alpha_{i}$ and $\theta_{i}$ 
and that $\theta_{i}$ does not depend on $\Delta_{M}$.
the prior $\prior \left( \alpha | \Delta_{M}, \theta \right)$ describes our model for the population of BM signatures. 
The prior on the event SM parameters $\theta_{i}$ is the 
posterior obtained from fitting the SM waveform.

Note that we marginalize over all the SM parameters, since they
may all affect the residual and the templates used to compute the $\alpha$ parameters.
However, not all SM parameters may enter our hierarchical models. Only a subset $T \in \theta$ may appear in $\prior (\alpha_{i} | \theta_{i} \Delta_{M})$ (for example, in Sec.~\ref{sec:Toy model} $T=\Mc$).
In this case, we may set the other parameters,  to their MLE value and only marginalize over $T$. 
This is computationally efficient but introduces a bias that scales as the inverse of the SNR squared
(in the limit of large SNR).

\section{Template source parameter dependence}%
\label{app:Template source parameter dependence}

In this appendix, we argue that for small signals and a reasonable choice of
template, the hyperparameters describing the dependence of the BM signature on the source
parameters can be recovered from the BM SNR regardless of the choice of template. 
This is important for agnostic searches for BM signatures of unknown form but whose
dependence on the source parameters is modeled.

Consider a CRPS signal $S(t;\theta, \Delta_{M})$ and a template $Z(t;\theta)$. 
We have used the notation to show that the dependence of $S$ on the source
parameters $\theta$ is determined by the hyperparameters $\Delta_{M}$.
If we can factor $S$ and $Z$ into a time-dependent and source parameter-dependent part, such that
 \begin{align}
	 \label{eq:factorization}
	 D \left( t ; \theta \right)
	 = 
	 D^{\theta} \left( \theta \right)
	 D^{t} \left( t \right)
	 , 
	 \quad 
	 Z
	 =
	 Z^{\theta} \left( \theta \right)
	 Z^{t} \left( t \right)
	 ,
\end{align}
then, the BM SNR also factorizes into a time-dependent and source parameter-dependent part:
\begin{align}
	\label{eq:SNR_dependence}
	\alpha ^{\theta}
	\alpha^{t}
	&=
	\left(
	Z^{\theta} \left( \theta \right)
	S^{\theta} \left( \theta \right)
	\right)
	\left(
	\int dt
	Z^{t} \left( t \right)
	S^{t} \left( t \right)
	\right)
\end{align}
and, as we know the template $Z(t;\theta)$, we can isolate $S( \theta )$.
The integral in $\alpha_{T}$ acts as a measure of the time match between the model
and the signal that is independent of the source parameters.

In our framework, the dependence of $S$ on the source parameters and BM hyperparameters
comes through a waveform-level parameter $\beta= \beta( \theta, \Delta_{M})$ (for example the 
tidal deformability introduced in Sec.~\ref{sec:Toy model}), which
is small when the BM signature is small.
This parameter is contained in $S^{\theta}$, and for small BM signatures we can write
\begin{align}
	\alpha
	\propto
	\left( 
		\beta^{2} 
		+ 
		\mathcal{O}\left(\beta^{3}\right) 
	\right),
\end{align}
where the leading order is $\theta^{2}$ since $\beta$ parametrised the waveform, which
is squared when computing the CRPS.

This property implies that we can recover the BM hyperparameters for any
reasonable choice of template, though selecting a template that closely matches
the true BM signature will lead to a higher SNR and more sensitive detection. 

The factorization of the signal into a time-dependent and source parameter-dependent part
relies on the assumption that the dependence of the template on the source parameters does
not vary greatly over the signal.
As a counter-example, consider choosing $Z(t;\theta) = \delta \left( t - \theta \right)$ or
$Z(t;\theta) = H \left( t - \theta \right)$, where $\delta$ is the Dirac delta function
and $H$ is the Heaviside step function. 
These choices may be interesting for studying excess power emitted at specific orbital separations,
or continuous losses due to dynamical friction, respectively.
In these cases, the template cannot be factorized into a time-dependent and
a source parameter dependent part. 
These templates should therefore not be used for agnostic searches, but are appropriate 
for specific searches for BM signatures with known time dependence.

\begin{figure}[htpb]
	\centering
	\includegraphics[width=\columnwidth]{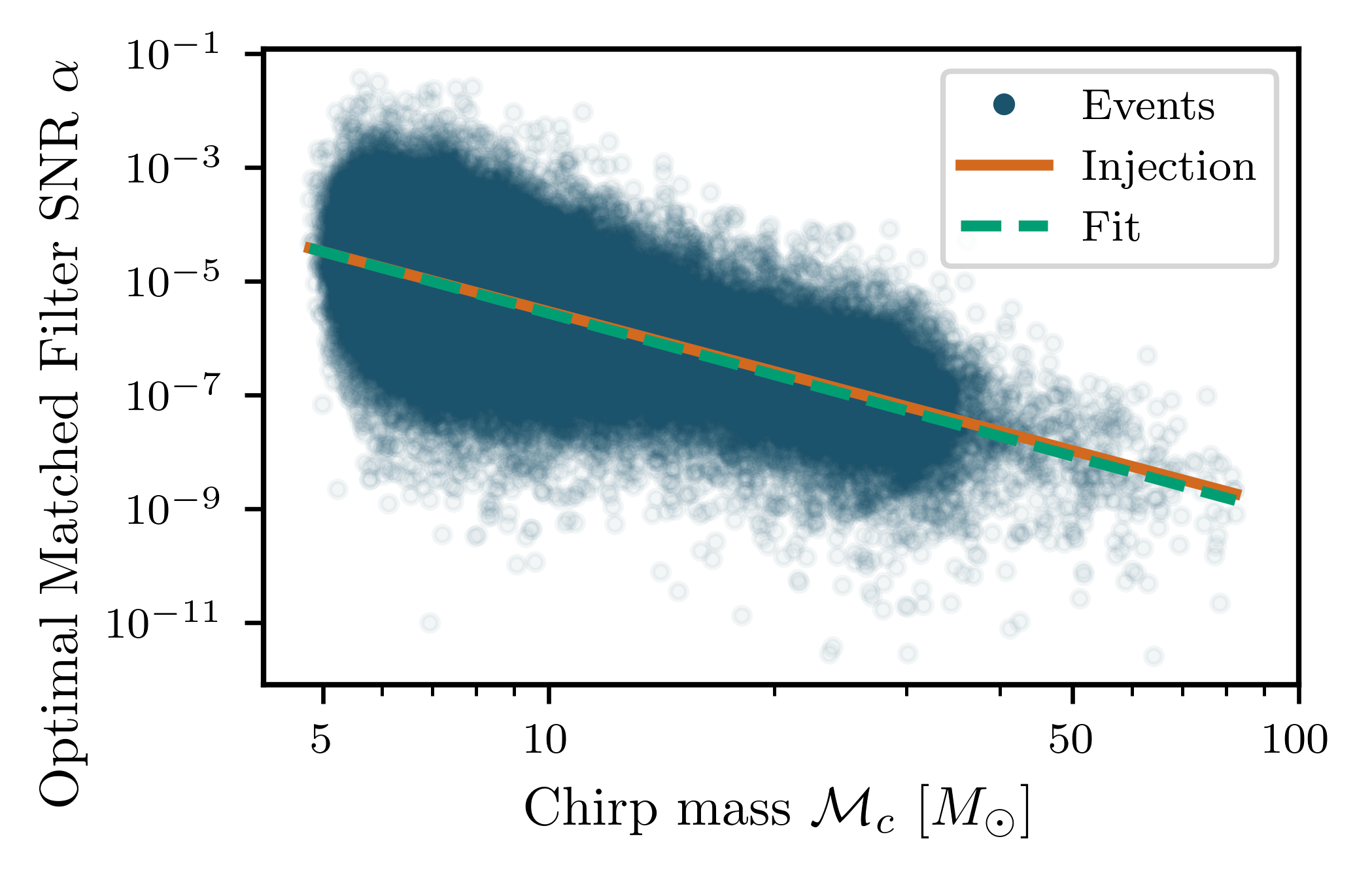}
	\label{fig:agnostic}
	\caption{BM SNR computed using the agnostic, orbital frequency-based template for the same events as in Fig.~\ref{fig:scaling}.}
\end{figure}

In Sec.~\ref{sec:Results}, we used $Z=S$ to compute the best-case BM SNR. To illustrate that
results are not dependent on this choice, we show, in Fig.~\ref{fig:agnostic},
the BM SNR computed using the agnostic template
\begin{align}
Z(t;\theta)
&=
\left \langle 
	\frac{d \ln f_{\detone} (\theta)}{d t}
	\frac{d \ln f_{\dettwo} (\theta)}{d t}
\right \rangle,
\end{align}
where $f_{\detone}$ and $f_{\dettwo}$ are the orbital frequencies computed using the SM
waveform model in detectors $\detone$ and $\dettwo$. 
The figure is comparable to the left panel of Fig.~\ref{fig:scaling} (where $D=3$ and $\beta_{0}=5$), 
and shows the BM SNR computed for the same events.
The same scaling with chirp mass is recovered, although there is a larger spread in the SNR recovered.
The recovered scale ($\alpha_{0}$) is smaller than when $Z=S$ since the template and signal are not perfectly
matched in time. 
The decrease is only from $3.6 \times 10^{-5}$ to $3.2 \times 10^{-5}$, as most of the SNR comes from the 
final stages of the inspiral where the template and signal are well matched.

\section{Population models} \label{sec:population_models}

We sample the source parameters used in our analysis from the following distribution:
\begin{align}
	P(\vec{\theta})
	&=
	P \left( m_{1},m_{2} \right) P \left( \chieff \right) 
	P(z) 
	P \left( \vec{\theta}_\text{ext}  \right) ,
\end{align}
where $\vec{\theta}_\text{ext}$ includes the sky location, the angle between the total angular momentum and the line of sight, the polarisation angle, geocentric time and GW phase at merger. 
Notably, we assume that all parameters are independent, except for the primary and secondary masses.
The samples are obtained from $P\left( \vec{\theta}_\text{GR}  \right)$ by inverse transform sampling.
We use uniform distributions for the extrinsic parameters and a Gaussian distribution for the effective spin 
$\chieff$ with mean $0.06$ and standard deviation $0.12$ as found in~\cite{abbottPopulationPropertiesCompact2021}.

The form and parameters used for the mass, redshift and spin distributions are described below and summarized in Table~\ref{tab:population_parameters}.

\subsection{Redshift evolution}

\begin{figure}[htpb]
	\centering
	\includegraphics[width=\columnwidth]{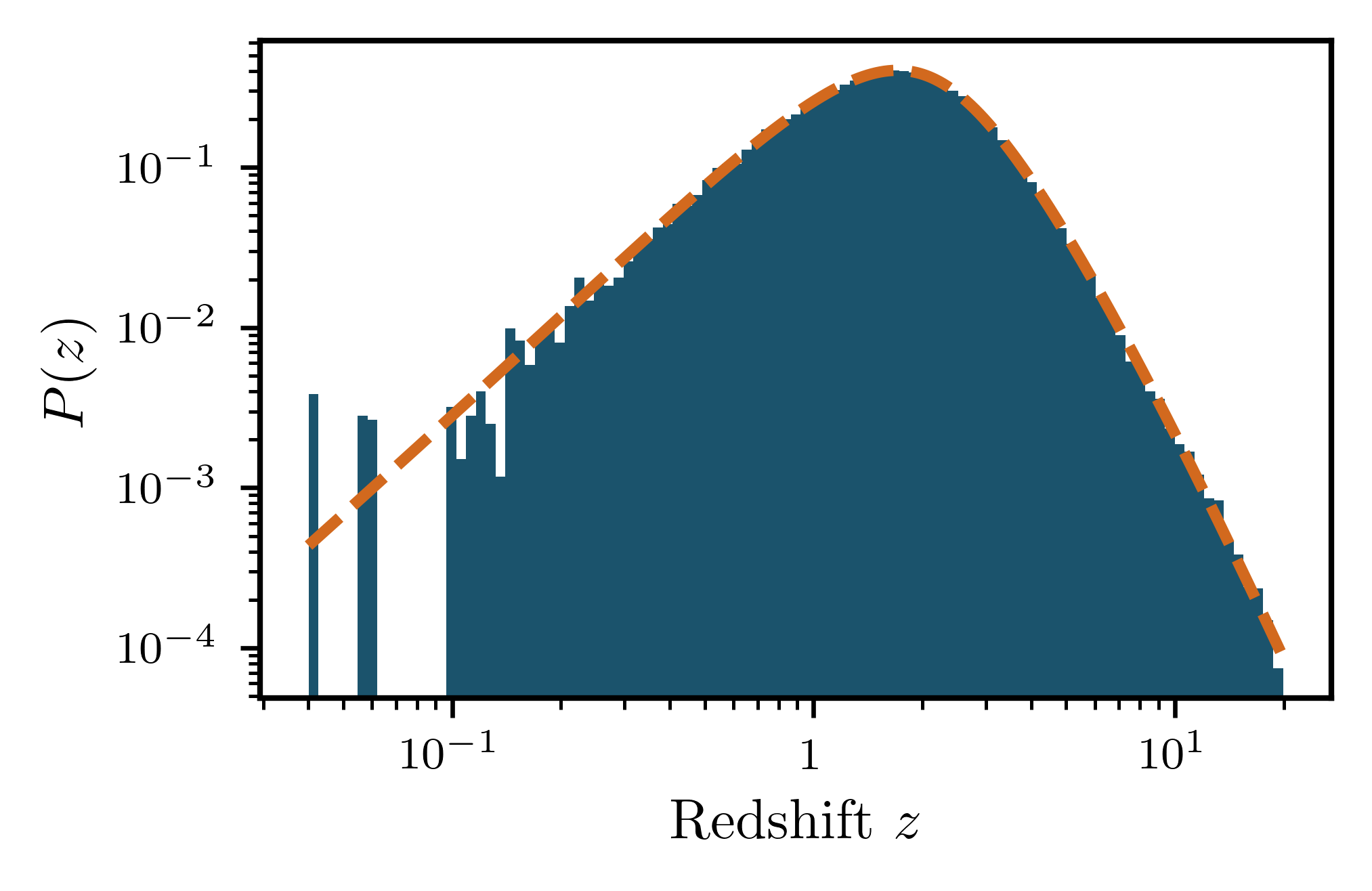}
	\caption{Samples drawn from the redshift distribution. The analytical expression is shown by the dashed line.}
	\label{fig:redshift}
\end{figure}

We assume that the evolution of the BBH merger rate per co-moving volume $V_{c}$ tracks the star formation rate up to some efficiency parameter.
For the Star formation rate, we use the fitting function used in~\cite{madauCosmicStarFormationHistory2014}, giving 
\begin{align}
	\frac{d N_\text{GW} }{d V_{c} dt_{s}}
	&\propto
	\left( 1 + z \right)^{\gamma}
	\frac
	{1 + \left( 1 + z_{p} \right)^{- \left( \gamma + \kappa \right) }}
	{1 + \frac{1 + z}{1 + z_{p}}^{\gamma + \kappa}},
\end{align}
with $\gamma=2.7$, $\kappa=2.9$ and $z_{p}=1.9$.
The rate in the source frame can be converted into the merger rate at the detector (per detector time $t_{d}$) per redshift.
\begin{align}
	\frac{d N_\text{GW} }{d z d t_{d} }
	&=
	\frac{1}{1+z}
	\frac{d V_{c}}{dz}
	\frac{d N_\text{GW} }{d z dt_{s}},
\end{align}
where the co-moving volume element over a shell of thickness $dz$ is
\begin{align}
	\frac{d V_{c}}{dz}
	&=
	4 \pi 
	\frac{c}{H_{0}}
	\frac{
		(1+z)^{2} D_{A}^{2}(z)
	}
	{E(z)},
\end{align}
where $E(z)=\sqrt{\left( 1+z \right)^{3} \Omega_{M} + \Omega_{\Lambda}}$ and $D_{A}(z)$ is the angular diameter distance.
We assume a flat $\Lambda$CDM cosmology, with $H_{0} = 70 \text{ kms}^{-1}\text{Mpc}^{-1}$, $\Omega_{\Lambda} = 0.7$ and $\Omega_{M}=0.3$. 
We show samples drawn from this distribution in Fig.~\ref{fig:redshift}, along with the analytical 
expression.

\subsection{Mass distribution}

\begin{figure}[htpb]
	\centering
	\includegraphics[width=\columnwidth]{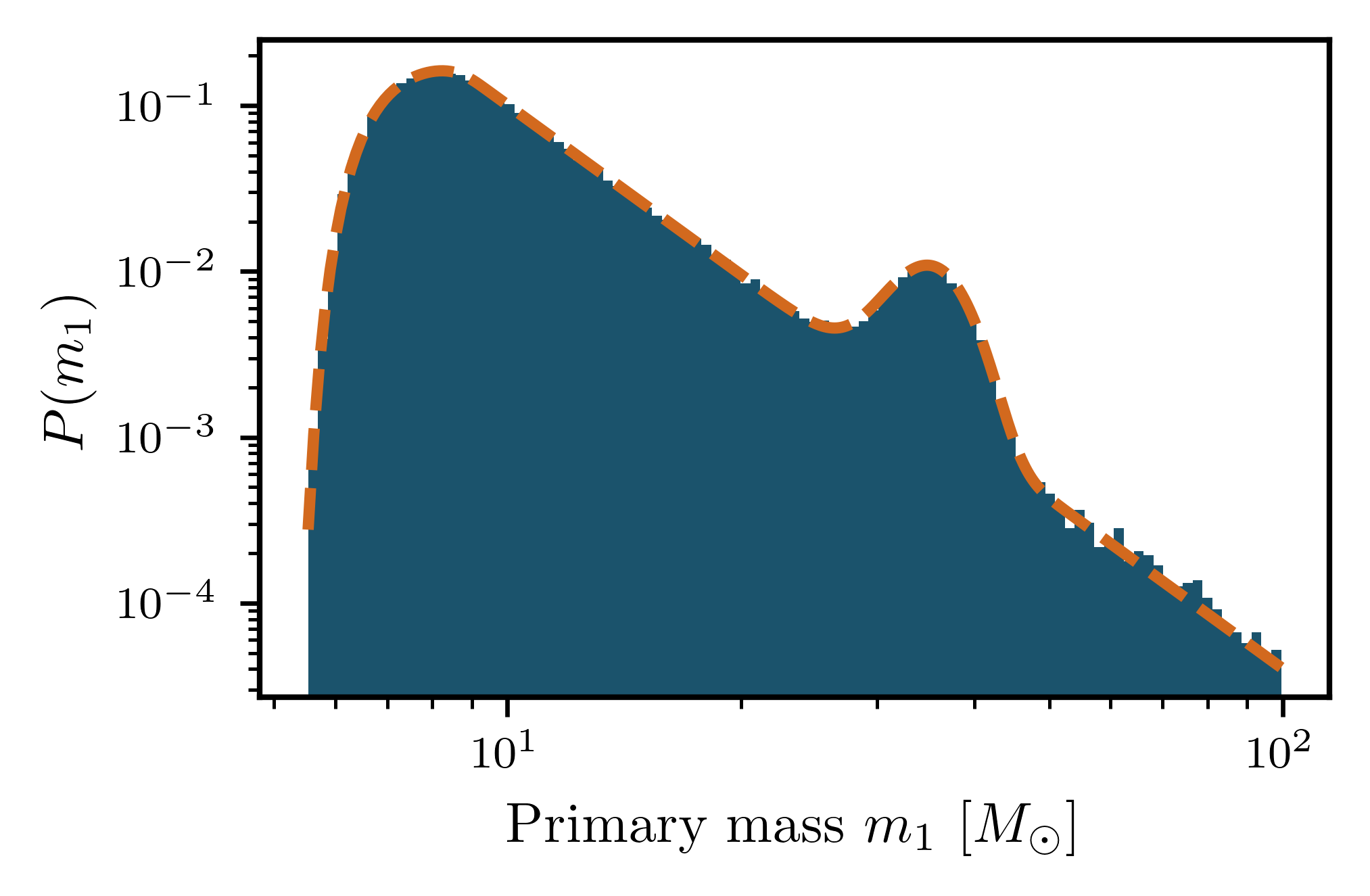}
	\caption{Samples drawn from the mass distribution. The analytical expression is shown by the dashed line.}
	\label{fig:mass}
\end{figure}

We use the Power law + Gaussian peak model from \cite{talbotMeasuringBinaryBlack2018} with the best-fit parameters from \cite{abbottPopulationPropertiesCompact2021,abbottPopulationMergingCompact2023}. 
This model gives the distribution of the larger, primary mass $m_{1}$ as a mixture of a power law and a Gaussian peak:
\begin{align}
	&P(m_{1}|\Lambda)\\
	\propto&
	\big[
		\left( 1 - \lambda \right) 
		P_\text{pow} (m_{1} | \alpha, m_\text{min} , m_\text{max} ) \\
	       &+
		\lambda
		P_\text{peak} \left( m_{1}|\mu_{m}, \sigma_{m}  \right) 
	 \big]
	S(m_{1},m_\text{min},\delta_{m}).
\end{align}
where $P_\text{pow}$ is a truncated power law with index $\alpha$ (we set it to 3.4) and support
between $m_\text{min}$ ($5M_{\odot}$) and $m_\text{max}$
($100M_{\odot}$), $\lambda$ (0.04) is the fraction of the population that comes from
the Gaussian peak, $P_\text{peak}$ is a Gaussian with mean at $\mu_{m}$
($35M_{\odot}$) and $\sigma_{m}$ ($3.9M_{\odot}$) standard deviation.
and $\sigma_{m}$ standard deviation.
A smoothing function $S$ is applied to avoid discontinuities at lower masses:
\begin{align}
	&S(x,m_\text{min},\delta_{m} )\\
	&=
	\left( 
		\exp 
		\left( 
			 \frac{\delta_{m}}{x - m_\text{min} }
			+
			\frac{\delta_{m}}{x - m_\text{min} - \delta_{m}}
		\right) 
		+
		1
	\right) ^{-1},
\end{align}
for $m_\text{min} \le m < m_\text{min} + \delta_{m}$, $0$ when $m < m_\text{min}$ and $1$ when $m \ge m_\text{min} + \delta_{m}$. We use $\delta_{m}=4.8$.
We show samples drawn from this distribution in Fig.~\ref{fig:mass}, along with the analytical expression.

The secondary mass is drawn assuming a power law in the mass ratio. This means that the conditional
probability from which $m_{2}$ is drawn is
\begin{align}
	&P(m_{2}|m_{1},m_\text{min}, \delta_{m})\\
	&\propto
	\begin{cases}
		\left( \frac{m_{2}}{m_{1}} \right) ^{\beta}
		S(m_{2},m_\text{min} ,\delta_{m}), 
		&
		m_{2}< m_{1},\\
		0, & \text{otherwise}.
	\end{cases}
\end{align}
Equal mass ratios are preferred, and we use $\beta=1.3$.

\section{Fisher Analysis}%
\label{app:Fisher analysis}

In the analysis carried out in Sec.~\ref{sub:Constraints for CE-like network}
with the hierarchical Bayesian model described in
Appendix~\ref{app:Hierarchical Bayesian Model}, we need to obtain the 
the likelihood $\likelihood \left( d_{i} | \alpha_{i}, \theta_{i} \right)$
on the SM parameters $\theta_{i}$ and the BM parameter $\alpha_{i}$ for each event $i$.
A full computation of this likelihood is expensive and not necessarily informative 
for a forecast, where we are interested in the general precision of the method.
Instead, we approximate it by a Gaussian distribution around the true values 
of $\alpha_{i}$ and $\theta_{i}$.
The values of the BM parameters $\theta_{i}$ are drawn from the population
models described in Appendix~\ref{sec:population_models} for each event $i$.
For our choice of template, $\alpha_{i}$ is the optimal match filtering SNR of the
CRPS template and is given by $\alpha = \langle Z | Z \rangle^{-\frac{1}{2}}$,
where 
\begin{align}
	\left \langle A | B \right \rangle
	&\equiv
	\int^{t_e}_{t_s} dt
	W(t)
	A(t)
	B(t).
\end{align}
The template function $Z$ we use is the most accurate CRPS template, and is therefore the
best-case scenario for recovery.

We expand the likelihood around the maximum likelihood estimate (MLE) of the parameters
and retain only the leading-order term, leaving us with a Gaussian distribution 
\begin{align}
	\label{eq:ln_likelihood_expansion}
	&\ln \mathcal{L}(d | \phi)
	\nonumber
	\\
	&=
	\ln \mathcal{L}(d_{i} | \phi_{a}^{\text{MLE}})
	\\
	+&
	\frac{1}{2}
	\left(
		\phi - \phi^{\text{MLE}}
	\right)^{a}
	\left(
		\phi - \phi^{\text{MLE}}
	\right)^{b}
	\left(
		\frac{\partial^{2} \ln \mathcal{L}}{\partial \phi^{a} \partial \phi^{b}}
	\right)\bigg|_{\phi_{\text{MLE}}}
	\\
	+&
	\mathcal{O}(\phi^{3}),
\end{align}
where $\phi = \{ \alpha, \theta \}$, which includes both the BM SNR and the source parameters.
Specific noise realizations will scatter the MLE around the true value of the parameters.
However, the expectation value of the MLE with respect to draws of the noise distribution
is the true value of the parameters. For a forecast, specific noise realizations are not
important, and we can set the MLE to its expectation value. 
In our analysis, the true values of the SM parameters are drawn from the population models
described in Appendix~\ref{sec:population_models}.
For the results in Sec.~\ref{sub:Constraints for CE-like network}, the true values of $\alpha$
are set to 0 for all events.

Assuming the likelihood is Gaussian (which is valid around the MLE in the limit of large SNR),
then the inverse of the Fisher
matrix,
\begin{align}
	\label{eq:fisher_matrix}
	\Gamma_{ab}
	&\equiv
	-
	E\left[
	\left(
		\frac{\partial^{2} \ln \mathcal{L}}{\partial \phi^{a} \partial \phi^{b}}
	\right)\bigg|_{\phi_{\text{MLE}}}
	\right],
\end{align}
where $E$ denotes the expectation value over the noise distribution, gives the
lower bound on the covariance of the MLE (the Cram\'er-Rao
bound~\cite{cramerMathematicalMethodsStatistics1991,raoInformationAccuracyAttainable1992a}).

In the \score framework, the likelihood $\likelihood \left( d_{i} | \phi_{i}
\right)$ for events labelled $i$ is computed using both the SM waveform model
on the strain data and the BM CRPS model on the residuals obtained. 
We can compute a Fisher matrix for each of these models, each matrix
a 15+$n$ dimensional square matrix, where $n$ is the number of BM parameters
(15 SM parameters and one BM parameter, $\alpha$, in our case).

The Fisher matrices for both measurements are combined to obtain 
the total Fisher matrix:
\begin{align}
	\label{eq:fisher_matrix_sum}
	\Gamma_{ab}
	&=
	\Gamma_{ab}^{\text{SM}}
	+
	\Gamma_{ab}^{\text{BM}}.
\end{align}
The total Fisher matrix is then inverted to obtain the covariance matrix of the 
likelihood $\likelihood \left( d_{i} | \phi_{i} \right)$.
In our study, however, the correlation between the SM and BM parameters is zero,
so the Fisher matrix is block diagonal and the likelihood is a product of two independent distributions.

For the SM model $m \left( t ; \theta \right)$, we assume Gaussian, stationary noise and high SNR.
With these assumptions, the Fisher matrix takes the simple form:
\begin{align}
	\label{eq:fisher_matrix_sm}
	\Gamma_{ab}^{\text{SM}}
	&=
	4
	\sum_{\text{det}}
	\Re
	\int_{0}^{\infty}
	df
	\frac{1}{S_{n} \left( f \right)}
	\left(
		\frac{\partial \tilde{m}}{\partial \phi^{a}}
	\right)^{\ast}
		\frac{\partial \tilde{m}}{\partial \phi^{b}}
\end{align}
where $S_{h} \left( f \right)$ is the one-sided power spectral density of the
strain noise and $\tilde{m}$ is the Fourier transform of the SM waveform model
$m \left( t ; \theta \right)$. 
The information obtained from each detector is summed over to obtain the
network Fisher matrix. Since $m \left( t ; \theta \right)$ does not depend on
the BM parameters ($\alpha$), the corresponding row and column in the Fisher
matrix are set to zero.

The likelihood for the CRPS signal $D^{IJ}_{i}$ is given by
\begin{align}
	\ln
	\likelihood
	\left( 
		D^{IJ} | \phi
	\right)
	&\propto
	\left\langle
	D^{IJ} \left( \theta \right) - Z ( \phi)
	\middle |
	D^{IJ} \left( \theta \right)-  Z (\phi)
	\right\rangle,
\end{align}
For simplicity, we assume that the value of $\theta$ used to compute $D^{IJ}$ are
the true values. 
This will be true in the limit of a small unmodelled signal.
With this assumption, the CRPS and template do not depend on the SM parameters.
Therefore, the only non-zero element of the Fisher matrix is $\Gamma^{\text{BM}}_{\alpha \alpha}$.
In the current study, we constructed the templates so that the morphology of
the template $Z$ only depends on the SM parameters, while its SNR for a given
event is given by the BM parameter $\alpha$.
We can then normalize the templates and $\Gamma^{\text{BM}}_{\alpha \alpha}=1$.

Since, in our model, $\alpha$ only depends on the chirp mass, we marginalize over 
the other SM parameters. This is simply removing the corresponding rows and columns
corresponding to the marginalized parameters in the covariance matrix.
The final likelihood is the product of two independent Gaussian distributions):
\begin{align}
	\likelihood \left( \alpha | \Mc \right)
	\propto
	&\exp
	\left(
		-
		\frac{1}{2}
		\left(
			\alpha - \alpha_{\text{T}} \left( \mathcal{M}_{c, \text{T}}  \right)
		\right)^{2}
	\right)
	\nonumber
	\\
	\times
	&\exp
	\left(
		-
		\frac{1}{2}
		\Gamma_{\Mc \Mc}^{\text{SM}}
		\left(
			\Mc - \mathcal{M}_{c,\text{T}}
		\right)^{2}
	\right).
\end{align}

To obtain the likelihood in $\Delta_{M} = \{ \alpha_{0}, D \}$, we must then
compute the integral 
\begin{align}
	\label{eq:mc_alpha_marginalisation}
	&\likelihood \left( \Delta_{M} | \{d_{i}\} \right)
	\nonumber	
	\\
	&=
	\int d \alpha
	d \Mc
	\likelihood \left( d_{i} | \alpha, \Mc \right)
	p \left( \alpha | \Mc, \Delta_{M} \right)
	\prior \left( \Mc \right),
\end{align}
which marginalizes over all possible mapping of $\alpha$ and $\Mc$ to
$\Delta_{M}$. We compute this integral numerically for each event.

\section{Mass range influence on inference} 
\label{app:Mass range influence on inference}

In this appendix, we derive the posterior computed to obtain Fig~\ref{fig:influence}, which shows the 
how well different chirp mass ranges constrain the BM hyperparameter $D$ variations
in the SNR and event distribution across chirp masses.
We assume a power law decay in the SNR with the chirp mass, so $\alpha(D,\Mc) =
\alpha_{0} \left( \Mc / 5M_{\odot} \right)^{\frac{5}{2}-2D}$, and the
Power-Law + Gaussian Peak mass distribution described in Appendix~\ref{sec:population_models}.
For simplicity, we assume: that the scale of the power law, $\alpha_{0}$, is known exactly; and that the
SM parameters (mass and others) are known exactly and are the same for all events so that the SNR 
only depends on the chirp mass;

The posterior on $D$ for $N$ events and for a set of data $\{d_{i}\}$ is given by:
 \begin{align}
	\label{eq:}
	\posterior(D| \{d_{i}\})
	&\propto
	\prior(D)
	\prod_{i}
	\likelihood(d_{i}|D).
\end{align}
If the chirp masses $\Mc$ of each event are known exactly, we can write:
\begin{align}
	\posterior(D| \{d_{i}\})
	&\propto
	\prior(D)
	\prod_{i}
	\likelihood \left( d_{i} | D, \Mc \right),\\
	&\text{where}\\
	\quad
	\likelihood \left( d | D, \Mc \right)
	&\propto
	\exp 
	\left( 
		-
		\frac{1}{2 \sigma_{\alpha}^{2}}
		\left(
			\alpha(D,\Mc)
			-
			\alpha(D_{T},\Mc)
		\right)^2	
	\right),
\end{align}
where $\sigma_{\alpha}=1$ is the standard deviation of the SNR , and $D_{T}$ is
the true value of $D$.
The likelihood can be exponentiated to give:
\begin{align}
	\label{eq:}
	\posterior(D| \{d_{i}\})
	&\propto
	\prior(D)
	\exp
	\left( 
		\sum^{N}_{i}
		\ln
		\likelihood
		(d_{i} | D, \Mc)
	\right).
\end{align}
In the limit of large numbers of events, we approximate the sum with the expectation value of the 
log-likelihood over the mass range:
\begin{align}
	\label{eq:Mc_posterior}
	&\posterior(D| \{d_{i}\})\\
	\propto&
	\prior(D)
	\exp
	\left( 
		N
		\int^{M_{\text{max}}}_{M_{\text{min}}}
		d \Mc
		P(\Mc)
		\ln
		\likelihood
		(d | D, \Mc)
	\right),
\end{align}
where $P(\Mc)$ is the mass distribution, including selection effects, and $M_{\text{min}}$ and $M_{\text{max}}$ are the minimum and maximum masses in the mass range.

\begin{figure}[htpb]
	\centering
	\includegraphics[width=\columnwidth]{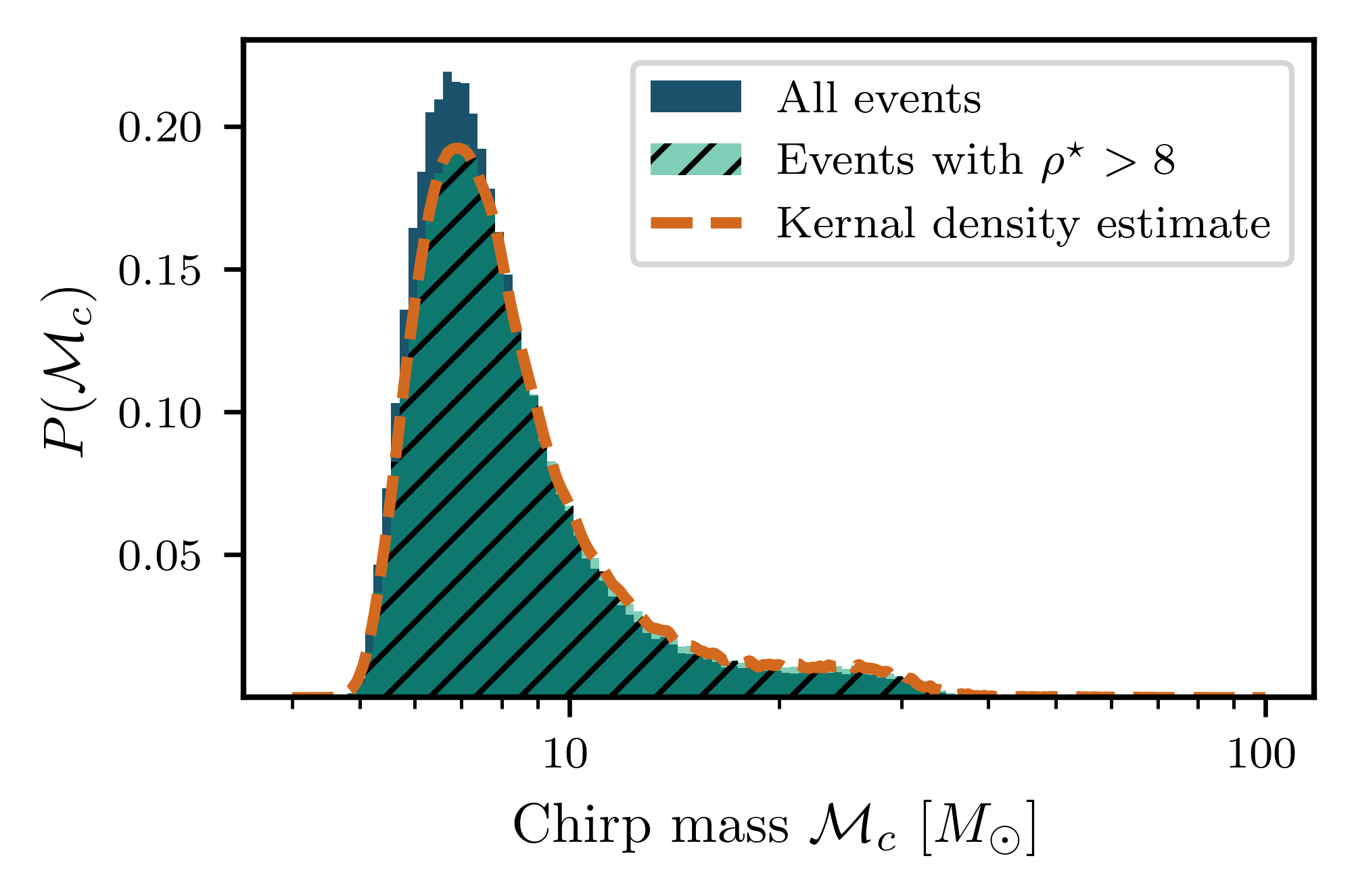}
	\caption{Kernel Density Estimation of the chirp mass distribution.
		The full distribution of events, as drawn from the population models 
		described in Appendix~\ref{sec:population_models}, is shown in solid blue.
		The distribution of events with match-filtering SNR above 8 is shown in hatched green.
		We use the Kernel Density Estimation in dashed orange to approximate the mass distribution
		$P(\Mc)$ in the posterior computation of Eq.~\eqref{eq:Mc_posterior}.
.}
	\label{kde_mc}
\end{figure}

We approximate $P(\Mc)$ using Kernel Density Estimation on the samples drawn from the full catalog described 
in Appendix~\ref{sec:population_models} and discarding samples with match-filtering SNR below 8. 
In Fig.~\ref{kde_mc}, we show the full event distribution as a function of
chirp mass in dark, solid blue; the selected event distribution with the
hatched green, and the Kernel Density Estimation in dashed orange.

\end{document}